\documentclass[12pt]{article}
\usepackage{latexsym}
\usepackage{fontenc}
\usepackage{amssymb}
\usepackage{amsmath}
\usepackage[all]{xy}
\usepackage[dvipdf]{epsfig}
\usepackage{color}
\usepackage{graphicx}
\usepackage{amsfonts}
\newcommand{\EA}[1]{\Gamma_{#1}}
\newcommand{\BA}[1]{S_{#1}}
\newcommand{\vol}{\int\!\!\textrm{d}^dx\,}
\numberwithin{equation}{section}

\begin{document}
\title{\begin{flushright}
\normalsize{MZ-TH/09-16}
\bigskip
%\vspace{1cm}
\end{flushright}
Bare vs. Effective Fixed Point Action in Asymptotic Safety: 
The Reconstruction Problem\footnote{Talk given by M.R. at the {\it Workshop on Continuum and Lattice Approaches
 to Quantum Gravity,} September 2008, Brighton, UK.}}
\date{}
\author{Elisa Manrique 
and Martin Reuter\\
Institute of Physics, University of Mainz\\
 Staudingerweg 7, D-55099 Mainz, Germany}
\maketitle

\begin{abstract} 
We propose a method for  the (re)-construction of a regularized functional integral,
well defined in the ultraviolet limit, 
from a solution of the functional renormalization group equation of 
 the effective average action. The functional integral is required to reproduce 
this solution. The method is of particular interest for asymptotically safe theories.
The bare action for the Einstein-Hilbert truncation of Quantum Einstein Gravity (QEG)
is computed and its flow is analyzed.
As a second example conformally reduced gravity is explored. 
 Various conceptual issues related to the reconstruction problem are discussed.
 
\end{abstract}

\section{Introduction}\label{I}

One of the major challenges of contemporary theoretical physics is the search for a compelling quantum theory of gravity. Despite the great efforts made until now, the complete description of such a theory seems to be far from completion. However, there exists a variety of
approaches that could be enlightening for  understanding certain aspects of what might ultimately be the correct quantum theory of gravity
 \cite{kiefer,A,R,T}.  

Certainly, one of the most interesting points of view one can adopt arises from the observation that the failure of perturbative approaches in gravity does not imply that such a quantum theory cannot exist. In principle there is the possibility of  quantizing gravity non-perturbatively, with the aid of Exact Renormalization Group techniques, say.
In fact within the so called  asymptotic safety program \cite{wein}-\cite{livrev},
a lot of efforts were devoted to establishing the existence of an ultraviolet fixed point at which Quantum Einstein
 Gravity (QEG) can be renormalized. Detailed calculations revealed that the renormalization group (RG) flow of 
the theory  does indeed possess an appropriate non-Gaussian fixed point (NGFP) in all approximations which were investigated.

Formulating QEG in terms of the gravitational average action  as proposed in \cite{mr},
the RG flow in question is that of the effective average action $\EA{k}[g_{\mu\nu},\cdots]$, henceforth abbreviated EAA \cite{avact},\cite{ymrev}.
While similar in spirit to the idea of a Wilson-Kadanoff renormalization, it replaces the iterated coarse graining 
procedure 
by a direct mode cutoff at the infrared (IR) scale $k$. More importantly,
the EAA is a scale dependent version of the ordinary { \it{effective}} action, while a ``genuine'' Wilsonian action 
$\BA{\Lambda}^{W}$ is a {\it{bare}} action, i.e. it is to be used under a regularized path integral.
As a result, it depends on
 the ultraviolet (UV) cutoff $\Lambda$; its  dependence on $\Lambda$ 
 is governed by a RG equation which is different from that for $\EA{k}$.
In fact the scale dependence of $\EA{k}$ is governed by a functional RG equation (FRGE)
which is one of the most useful items in the EAA ``tool box".

As a quantization method, the FRGE is in principle sufficient to  fully define a quantum field theory: given a complete RG 
trajectory, well defined for all values of $k\in [0,\infty)$, we have complete knowledge of all
properties of the QFT at hand.
Its Green's functions are the derivatives of 
 $\EA{k}$ and at $k=0$ they  coincide with those of the standard effective action $\EA{}\equiv\EA{k=0}$ \cite{avact}.
 The RG trajectory chosen must be free from divergences in both
the IR and the UV limit.To realize the asymptotic safe property, the trajectory should be arranged
to hit the NGFP  in the UV limit.

However, one should stress the difference between the EAA and the Wilsonian approach.
In a sense, $\BA{\Lambda}^{W}$ for different values of $\Lambda$ is a set of actions for the same system: 
the Green's functions have to be computed from $\BA{\Lambda}^{W}$ by a 
further functional integration over the low momentum modes, and this integration renders them  independent of $\Lambda$. 
By contrast, the EAA can be thought of as the standard effective action for a family of different systems: for any
 value of $k$ it equals the standard effective action of a model with the bare action $\BA{\Lambda}+ \Delta_{k}S$
 where $\Delta_{k}S$ is the mode suppression term. The corresponding $n$-point functions, computed
as functional derivatives of $\EA{k}$ without any further integration, are scale dependent and 
they provide an effective field theory description \cite{bh}-\cite{mof}
of the physics at scale $k$. 
%(See \cite{avactrev,livrev} for a disscussion of this point.)

Because of these differences between the EAA,  $\EA{k}$,  and a genuine Wilson action $\BA{\Lambda}$, this way of constructing 
an asymptotically safe field theory does not by itself yield  a regularized path integral over metrics $\gamma_{\mu\nu}$ whose
continuum limit would be related to the  RG trajectory $\{\Gamma_k, 0\leq k<\infty\}$ in a straightforward way.

In general the relationship between $\EA{k}$ and $\BA{\Lambda}$ will depend on how we regularize the path integral measure 
$\mathcal{D}_{\Lambda}\gamma_{\mu\nu}$ when defining the generating functional.
In the following we shall demonstrate that it is possible to reconstruct a regularized functional integral such that it describes a fixed,
prescribed asymptotically safe theory in the infinite cutoff limit $\Lambda\rightarrow \infty$.
Adopting a particularly convenient UV regularization scheme we shall see that the information 
contained in $\EA{k}$ is sufficient in order to determine the related bare action $\BA{\Lambda}$ in the limit 
$\Lambda\rightarrow\infty$.
Given a complete RG trajectory $\{\EA{k}, 0\leq k<\infty\}$, computed from a FRGE {\it without any UV regulator}, 
we deduce from it how the bare coupling constants contained in 
$\BA{\Lambda}$ must behave in the UV limit when the path integral (with the measure $\mathcal{D}_{\Lambda}\gamma$ and 
action $\BA{\Lambda}$ defined according to the special scheme adopted) is required to reproduce the prescribed $\EA{k}$ 
trajectory. 

There are various motivations for trying to construct a path integral representation of asymptotically safe QEG: 

{\bf (a)} Working with the EAA alone we have no  access to  the microscopic (or ``classical'') system whose standard
quantization gives rise to this particular effective action. A functional integral
representation of the asymptotically safe theory will allow the reconstruction of the microscopic degrees 
of freedom that we implicitly  integrated out in solving the FRGE, as well as their fundamental
interactions. The path integral provides us with their action, and from this action, by a kind of generalized
Legendre transformation, we can reconstruct their Hamiltonian description. From this phase space 
formulation we can read off the classical system whose quantization (also by other methods, canonically say)
leads to the given effective action. We expect this system to be rather complicated so that it cannot be 
guessed easily. This is why we start at the effective level where we know what to look for, namely
a $\Gamma$ whose functional derivatives ($S$-matrix elements) are such that observable quantities have no divergences
on all momentum scales.

{\bf (b)} Many general properties of a quantum field theory are most easily analyzed in a path integral setting,
the implementation of symmetries, the derivation of Ward identities or the incorporation of constraints, to mention just a few. 

{\bf (c)} Many approximation schemes (perturbation theory, large-N expansion, etc.) are more naturally described in a path integral
 rather than a FRGE language. A standard way of doing perturbation theory is to compute, order by order, the counter terms to be included in
  $\BA{\Lambda}$ to get finite physical results in the limit $\Lambda\rightarrow\infty$. Now, QEG
is not renormalizable in perturbation theory and hence new counter terms with free coefficients must be introduced at each order.
If, on the other hand, QEG is asymptotically safe,
defined by a complete trajectory $\{\EA{k}, 0\leq k<\infty\}$, this trajectory ``knows" 
 the correct UV completion of the perturbative calculation. But in order to 
extract this information from $\Gamma_k$ and make contact with the
perturbative language of $S_{\Lambda}$-counter terms  we must convert the $\EA{k}$-trajectory to a $\BA{\Lambda}$-trajectory
first.

{\bf (d)} Ultimately we would like to understand how QEG relates to other approaches to quantum gravity, such as 
canonical quantization, loop quantum gravity \cite{A,R,T} or Monte Carlo simulations \cite{hamber}-\cite{ajl3}, in which the bare action often
plays a central role. In the Monte Carlo simulation of the Regge and dynamical triangulations formulation, for instance,
the starting point is a regularized path integral involving some discrete version of $\BA{\Lambda}$, and in order to take
the continuum limit one  must fine tune the bare parameters in $\BA{\Lambda}$ in a suitable way.
If one is interested in the asymptotic scaling, for instance, and wants to compare the analytic QEG predictions to the way
the continuum is approached in the simulations, one should convert the $\EA{k}$-trajectory to a $\BA{\Lambda}$-trajectory first.
The map from $\Gamma_k$ to $S_{\Lambda}$ depends explicitly on how  the path integral is discretized; 
so each alternative formulation of QEG has its own $S_{\Lambda}$ for one and the same $\EA{k}$.

The remaining sections of this paper are organized as follows. In Section \ref{II} we describe some features of the EAA when 
it has an additional UV cutoff built into it. Then, in Section \ref{reconstructing}, we explain the reconstruction of the bare 
from the running effective action. Thereafter the method is applied, in Section \ref{QEGtruncation}, to the Einstein-Hilbert truncation
of QEG, and in Section \ref{creh}, to conformally reduced gravity in the local potential approximation. A summary and outlook 
is given in Section \ref{sectionsix}.
In an appendix we further elaborate on the relation between the bare and the effective average action by means of a simple example
which is of physical interest in its own right: the cosmological constant induced by a scalar matter field.

%***********************************************
%**********************************************

\section{Effective Average Action with UV cutoff}\label{II}

In this section we describe how the functional integral underlying the definition of the effective average action  can be made well defined. 
We regularize it by introducing an UV cutoff $\Lambda$ and then derive, in a completely well defined way, the corresponding 
EAA and its flow equation in presence of $\Lambda$. Many different regularization schemes are conceiveable here.
For concreteness we use a kind of ``finite mode regularization"
 which is
ideally suited for implementing the ``background independence" mandatory in QEG.

For simplicity, we consider a single scalar field on flat space. The
generalization to more complicated theories can be achieved  by obvious notational changes.

Let $\chi(x)$ be a real scalar field on a flat $d$-dimensional Euclidean spacetime. 
In order to discretize momentum space 
we compactify spacetime to a $d$-torus. As a result, the eigenfunctions of the Laplacian 
$\Box=\delta^{\mu\nu}\partial_{\mu}\partial_{\nu}\equiv -\hat{p}^2$ are plane waves $u(x)\propto\exp{(ip\cdot x)}$
with  discrete momenta $p_{\mu}$ and eigenvalues $-p^2$. Given a UV cutoff scale $\Lambda$, there are only finitely
many eigenfunctions with  $|p|\equiv \sqrt{p^2}\leq\Lambda$. We regularize the path integral in the 
UV by restricting the integration to those modes.
Therefore, the field $\chi$ and the source $J$ have an expansion
\begin{align}
\chi(x)=\sum_{|p|\in[0,\Lambda]}\chi_p\; u_p(x),& \quad\textrm{and}\quad J(x)=\sum_{|p|\in[0,\Lambda]}J_p\; u_p(x)
\end{align}
Now we define a UV-regulated analogue of the standard functional $W_k[J]$:
\begin{equation}\label{wkl}
\exp{\Big(W_{k,\Lambda}[J]\Big)}\equiv \int\mathcal{D}_{\Lambda}\chi\exp\Big( -\BA{\Lambda}[\chi]-\Delta_k S[\chi]
+\vol J(x)\chi(x)\Big)
\end{equation}
The notation in eq.(\ref{wkl}) is symbolic. In fact, its RHS involves only finitely many integrations and is not a genuine
functional integral. 
Here, the measure $\mathcal{D}_{\Lambda}\chi$ stands for an integration over the Fourier coefficients $\chi_p$ with $p^2$
below $\Lambda^2$:
\begin{equation}\label{measure}
\int\mathcal{D}_{\Lambda}\chi=\prod_{|p|\in[0,\Lambda]}\int_{-\infty}^{\infty}\!\!\textrm{d}\chi_{p}\;M^{-[\chi_{p}]}
\end{equation}
The arbitrary mass parameter $M$ was introduced in order to give the canonical dimension zero to (\ref{measure}).
As always in the EAA construction \cite{avact,avactrev}, the IR modes with $|p|<k$ are suppressed 
by a IR cutoff $ \mathcal{R}_k(\hat{p}^2)$ which gives rise to a momentum dependent mass term:
\begin{equation}\label{IR}
\Delta_k S[\chi]=\frac{1}{2}\vol \chi(x)\mathcal{R}_k(\hat{p}^2)\chi(x)
\end{equation}
In (\ref{wkl}) the bare action $\BA{\Lambda}$ is allowed to depend on the UV cutoff. Ultimately we would  like to fix
this $\Lambda$-dependence in such a way that, for every finite $k$ and $J$, the path integral has a well defined limit for
$\Lambda\rightarrow\infty$.

Following the standard construction \cite{avact}, we define the EAA as
\begin{equation}\label{effectiveaction}
\EA{k,\Lambda}[\phi]\equiv \widetilde{\Gamma}_{k,\Lambda}[\phi]-\frac{1}{2}\vol \phi(x)\mathcal{R}_{k}(\hat{p}^{2})\phi(x)
\end{equation}
 where  $\widetilde{\Gamma}_{k,\Lambda}[\phi]$ is  the Legendre transform of $W_{k,\Lambda}[J]$ with respect to $J$ and
 $\phi=\{\phi_{p}\}_{|p|\in[0,\Lambda]}$ is the expectation value field $\phi(x)\equiv \langle \chi(x)\rangle$ obtained by 
differentiating $W_{k,\Lambda}[J]$ with respect to the source $J(x)$.

It is then straightforward to show that the definition (\ref{effectiveaction}) implies the following
exact FRGE for $\EA{k,\Lambda}$:
\begin{equation}\label{frge}
k\partial_{k}\EA{k,\Lambda}[\phi]=\frac{1}{2}\textrm{Tr}_{\Lambda}\Big[ \Big(\EA{k,\Lambda}^{(2)}[\phi]+\mathcal{R}_{k}\Big)^{-1}k\partial_{k}\mathcal{R}_{k}\Big]
\end{equation}
Here, $\textrm{Tr}_{\Lambda}$ denotes the trace restricted to the subspace spanned by the eigenfunctions of $p^{2}$ with eigenvalues smaller than $\Lambda^{2}$:
\begin{equation}\label{trace}
\textrm{Tr}_{\Lambda}[\cdots]=\textrm{Tr}\Big[\theta(\Lambda^{2}-\hat{p}^{2})[\cdots]\Big]
\end{equation}
It is worth mentioning that $\EA{k}$ satisfies the integro-differential equation
\begin{eqnarray}\label{integroeq}
\exp{\Big(-\EA{k,\Lambda}[\phi]\Big)}& = &\int\!\!\mathcal{D}_{\Lambda}f \exp{\Big(-\BA{\Lambda}[\phi+f]+    \vol f(x) \frac{\delta\EA{k,\Lambda}[\phi]}{\delta \phi(x)}}
\nonumber\\
& & {}-\frac{1}{2}\vol f(x)\mathcal{R}_{k}(\hat{p}^{2})f(x)\Big)
\end{eqnarray}
where we have introduced the fluctuation field $f(x)\equiv \chi(x)-\phi(x)$. Eq.\eqref{integroeq}
is the starting point for our investigations in the next section. 
%

%\subsection{Removing the UV cutoff from the FRGE}\label{removingUV}

A natural question that arises immediately is whether  the UV cutoff can be removed from the FRGE. Indeed,
for this to be possible it is  sufficient to assume that the cutoff is chosen such that $k\partial_k\mathcal{R}_{k}(p^{2})$ 
decreases rapidly enough so that the trace of the flow equation \eqref{frge} exists even in the 
limit  $\Lambda\rightarrow \infty$. As a result, the``$\Lambda$-free" FRGE without UV cutoff, valid for all $k\geq 0$, 
has the familiar form:

\begin{equation}\label{fluxeq1}
k\partial_k\EA{k}[\phi]=\frac{1}{2}\textrm{Tr}\Big[ \Big(\EA{k}^{(2)}[\phi] + 
\mathcal{R}_{k}(p^{2})\Big)^{-1}k\partial_k\mathcal{R}_{k}(p^{2})\Big]
\end{equation}
We denote the solutions  of \eqref{fluxeq1} as  $\{\EA{k},\; 0\leq k <\infty\}$,
and those of the FRGE \eqref{frge} with UV cutoff as  $\{\EA{k,\Lambda},\; 0\leq k <\Lambda\}$.

It is easy to show \cite{elisa1} that the flow equations for $\EA{k}$ and $\EA{k,\Lambda}$ are essentially the same
as long as $k\ll \Lambda$. Generically, when $k$ approaches $\Lambda$ from below, there exist 
some small corrections due to the UV cutoff which  affect $\EA{k,\Lambda}$ and cause it to differ from $\Gamma_k$. 
However, it is always possible to chose a special  IR cutoff
$\mathcal{R}_k(p^2) $ such that those corrections vanish. In particular, this happens with the optimized cutoff \cite{opt} 
$\mathcal{R}_k(p^2)=(k^2-p^2)\theta(k^2-p^2)$.
As a result the functional $\EA{k,\Lambda}$ satisfies the same FRGE as $\EA{k}$, but is defined  in the interval $k\leq\Lambda$ only.  
For identical initial conditions,  a simple relation between the solutions of the two flow equations exists therefore: 
\begin{equation}\label{inicond0}
\EA{k,\Lambda}=\EA{k}\quad\textrm{ when }0\leq k \leq\Lambda
\end{equation}
Here $\Lambda$ is a fixed, but arbitrary finite scale. 
In other words, $\{\EA{k,\Lambda},\; 0\leq k <\Lambda\}$ is the restriction of the complete solution $\{\EA{k},\; 0\leq k <\infty\}$
to the interval $k<\Lambda$.
Thus, sending $\Lambda\rightarrow\infty$ in \eqref{inicond0} is  a trivial step. The situation is illustrated in Fig.1.
\begin{figure}\label{figure2}
\centering
\includegraphics[width=10cm,height=7cm]{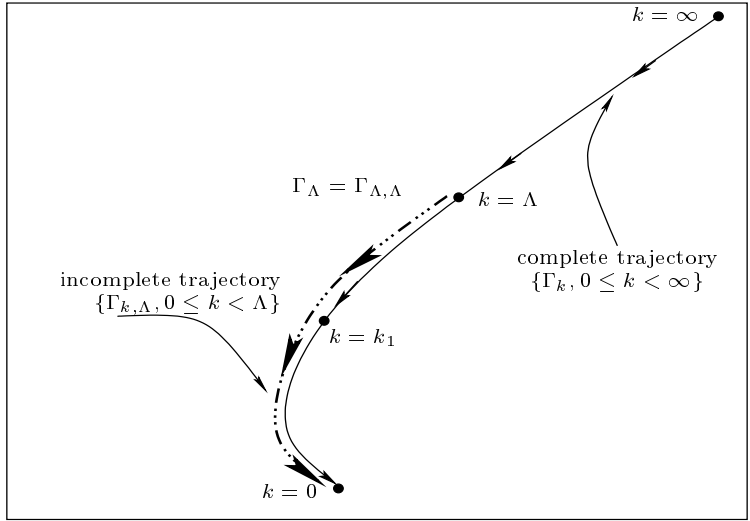} 
\caption{\small Employing the optimized cutoff every complete solution to the $\Lambda$-free FRGE gives rise to a  solution
of the FRGE with UV cutoff, valid up to any  value of $\Lambda$. }
\end{figure}

\section{Reconstructing the bare action}\label{reconstructing}

The problem we want to address now is how one can determine the $\Lambda$-dependence of the bare action
$\BA{\Lambda}$, given some complete solution of the $\Lambda$-free flow equation,  $\{\EA{k},\; k\in [0,\infty)\}$,
According to \eqref{inicond0}, this complete solution implies a solution with a UV cutoff: $\{\EA{k,\Lambda},\;  k\in [0,\Lambda]\}$.
Setting $k=\Lambda$ we have in particular $\EA{\Lambda,\Lambda}=\EA{\Lambda}$ or, more explicitly, 
\begin{equation}
\EA{k=\Lambda,\Lambda}=\EA{k=\Lambda}
\end{equation}
Thus, knowing $\EA{k}$ for all $k$ means that we know $\EA{\Lambda,\Lambda}$ for all $\Lambda$. 

Next we shall explain
how, given $\EA{k}$, the bare action $\BA{\Lambda}$ can be found.
In particular, setting  $k=\Lambda$ we are given $\EA{\Lambda,\Lambda}=\EA{\Lambda}$. Since $\EA{k}$ is a solution for all values of $k$
the action $\EA{\Lambda,\Lambda}$ is known for all values of $\Lambda$. 
%

%\subsection{The saddle point expansion}\label{sectionsaddle}

Using equation \eqref{integroeq} we can obtain the desired relation between $\EA{k}$ and $S_{\Lambda}$. 
For this purpose we evaluate the functional integral on the RHS of \eqref{integroeq} by a saddle point expansion.
Let the fluctuation field  be $f(x)\equiv f_{0}(x)+h(x)$ where $f_{0}$ is the stationary point of the action 
\begin{equation}
\BA{\textrm{tot}}[f;\phi]\equiv \BA{\Lambda}[\phi+f] -  \vol f(x) \frac{\delta\EA{k,\Lambda}[\phi]}{\delta \phi(x)}-\frac{1}{2}\vol f(x)\mathcal{R}_{k}(\hat{p}^{2})f(x)
\end{equation}
Now, expanding $S_{\textrm{tot}}$ to second order in $h$ and 
performing the Gaussian integral over $h$ we obtain the following relationship between the bare and the average action:
\begin{eqnarray}\label{loopexpansion}
\EA{k,\Lambda}[\phi] &=& S_{\Lambda}[\phi+f_{0}]-\vol f_{0}\frac{\delta\EA{k,\Lambda}[\phi]}{\delta\phi}+\frac{1}{2}\vol f_{0}\mathcal{R}_{k}f_{0}+{}
\nonumber\\
& & {}+  \frac{1}{2}\textrm{Tr}_{\Lambda}\ln{\Big[\Big(\frac{\delta^{2}S_{\Lambda}[\phi+f_{0}]}{\delta\phi^{2}}+\mathcal{R}_{k}\Big)M^{-2}\Big]}+\cdots
\end{eqnarray} 
Recalling that the stationary point $f_{0}$  has an expansion in powers 
of $\hbar$ too, \eqref{loopexpansion} yields, in a symbolic notation,
\begin{eqnarray}\label{rel2}
\EA{k,\Lambda}[\phi]-S_{\Lambda}[\phi] &=& -\int f_{0}\frac{\delta}{\delta\phi}\Big(\EA{k,\Lambda}-S_{\Lambda}\Big)[\phi] + 
\frac{1}{2}\int f_{0}\Big( S^{(2)}_{\Lambda}[\phi]+\mathcal{R}_{k}\Big)f_{0} + \mathcal{O}(f_{0}^{3})+{}
\nonumber\\
 & & {} \!\!\!\!\!\! \!\!\!\!\!\!\!\!\! \!\!\! \!\!\!\!\!\! \!\!\!\!\!\!\!\!\! \!\!\! + \frac{\hbar}{2}\;\textrm{Tr}_{\Lambda}\ln{\Big\{  \big[  S^{(2)}_{\Lambda}[\phi]+S^{(3)}_{\Lambda}[\phi]f_{0}+S^{(4)}_{\Lambda}[\phi]f_{0}f_{0}+
 \cdots+\mathcal{R}_{k}\big]M^{-2}\Big\}}+ \mathcal{O}(\hbar^{2})
\end{eqnarray}
Together with the expansion of the stationary point condition $(\delta S_{\textrm{tot}}/\delta f)[f_{0}]=0$ 
the above equation is
solved self-consistently if $f_{0}=0+\mathcal{O}(\hbar)$ and $\EA{k,\Lambda}[\phi]-S_{\Lambda}[\phi]=\mathcal{O}(\hbar)$,
which leads to the following 1-loop formula for the difference between the average and the bare action:
\begin{equation}
\EA{k,\Lambda}[\phi]-S_{\Lambda}[\phi]=\frac{1}{2}\textrm{Tr}_{\Lambda}\ln{\Big\{ \big[ S^{(2)}_{\Lambda}[\phi]+\mathcal{R}_{k} \big]M^{-2} \Big\} }
\end{equation}
Setting $k=\Lambda$ we arrive at the final result
\begin{equation}\label{finalequation}
\EA{\Lambda,\Lambda}[\phi]-S_{\Lambda}[\phi]=\frac{1}{2}\textrm{Tr}_{\Lambda}\ln{\Big\{ \big[ S^{(2)}_{\Lambda}[\phi]+\mathcal{R}_{\Lambda} \big]M^{-2} \Big\} }
\end{equation}
This is an equation to be solved for $\BA{\Lambda}$. It tells us how the bare action $\BA{\Lambda}$ must depend on $\Lambda$ in order
to give rise to the prescribed $\EA{\Lambda,\Lambda}$.
The relation (\ref{finalequation}), and its obvious generalizations to more complicated theories, is our main tool for (re)constructing the path integral
that belongs to a known solution of the FRGE. 

An important comment is in order here. Even though the parameter $M$ was introduced
in \eqref{measure} only in order  to make the measure dimensionless, it has a nontrivial impact on the
solution of \eqref{finalequation} for the bare action $\BA{\Lambda}$. Indeed, different
choices of $M$ can lead to quite different actions , but all of them are physically equivalent. (See \cite{elisa1} for a detailed
discussion.) In this sense, changing $M$ simply
amounts to shifting the contributions from the measure into the bare action. Therefore
neither $\int\mathcal{D}_{\Lambda}\chi$ nor $\exp{(-\BA{\Lambda})}$ have a physical meaning
separately, only the combination of them has.

\section{ QEG and the Einstein-Hilbert truncation}\label{QEGtruncation}

The results derived above can be  generalized to the case of Quantum Einstein Gravity \cite{elisa1}, 
following the  strategy for constructing the EAA as in \cite{mr} and implementing an UV cutoff in addition.
 Indeed, we shall use the same notations and conventions as in \cite{mr}to which the reader is referred 
for further details.

\subsection{The gravitational average action}\label{covariantUV} 

The construction of the gravitational average actions starts out from a path integral 
$\int\mathcal{D}\gamma_{\mu\nu}\exp{(-S[\gamma_{\mu\nu}])}$.
First we introduce a background metric $\bar{g}_{\mu\nu}(x)$, decompose the integration variable as 
$\gamma_{\mu\nu}\equiv \bar{g}_{\mu\nu}+ h_{\mu\nu}$, and gauge-fix the resulting path integral over $h_{\mu\nu}$.
It is this integral that we make well defined by introducing an UV cutoff into the measure along with an IR-suppression
term $\Delta_kS$ analogous to (\ref{IR}):
\begin{equation}\label{qegpath}
\int\mathcal{D}_{\Lambda}h\;\mathcal{D}_{\Lambda}C\;\mathcal{D}_{\Lambda}\bar{C}\;\exp{\Big(-\widetilde{S}_{\Lambda}[h,C,\bar{C};\bar{g}]
-\Delta_kS[h,C,\bar{C};\bar{g}]\Big)}
\end{equation} 
Here $C^{\mu}$ and $\bar{C}_{\mu}$ are the Fadeev-Popov ghosts, and the total bare action,
$
\widetilde{S}_{\Lambda}\equiv\BA{\Lambda}+\BA{\textrm{gf},\Lambda}+\BA{\textrm{gh},\Lambda}
$,
which is allowed to depend on $\Lambda$, includes the gauge fixing term $\BA{\textrm{gf},\Lambda}$ and the ghost
action $\BA{\textrm{gh},\Lambda}$.
The UV cutoff is implemented by restricting the expansion to eigenfunctions of the covariant Laplacian $\bar{D}^2\equiv \bar{g}^{\mu\nu}\bar{D}_{\mu}\bar{D}_{\nu}$ with eigenvalues $\kappa$ smaller than a given $\Lambda^2$.
Hence the measure reads in analogy with (\ref{measure})
\begin{equation}\label{qegmeasure}
\int\mathcal{D}_{\Lambda}h=\prod_{\kappa\in[0,\Lambda^2]}\prod_m\int_{-\infty}^{\infty}\!\!\textrm{d}h_{\kappa m}
\;M^{-[h_{\kappa m}]}
\end{equation}
and likewise for the ghosts. Here, $m$ is a degeneracy index.
 The remaining steps in the construction of the gravitational average action proceed exactly as in \cite{mr}.
Note that in this construction the background metric $\bar{g}_{\mu\nu}(x)$ is crucial not only for the gauge fixing and the 
IR cutoff, but also for implementing the UV cutoff.

The key properties of the functional thus defined are the exact FRGE and the integro-differential equation which it satisfies.
The flow equation reads
\begin{equation}\label{qegfloweq}
k\partial_k\EA{k,\Lambda}[\bar{h},\xi,\bar{\xi};\bar{g}]=\frac{1}{2}\textrm{STr}_{\Lambda}\Big[\Big(\EA{k,\Lambda}^{(2)}+\widehat{\mathcal{R}}_{k}\Big)^{-1}k\partial_{k}\widehat{\mathcal{R}}_{k}\Big]
\end{equation}
Here the supertrace ``STr" implies the extra minus sign in the ghost sector. In fact, the cutoff operator $\widehat{\mathcal{R}}_{k}$ and the Hessian $\EA{k,\Lambda}^{(2)}$ are matrices in the space of dynamical fields $\bar{h},\xi$ and $\bar{\xi}$.
The background covariant regularization of the measure entails the appearance of the restricted trace
\begin{equation}\label{qegtrace}
\textrm{STr}_{\Lambda}[\cdots]\equiv\textrm{STr}\Big[\theta(\Lambda^2+\bar{D}^2)[\cdots]\Big]
\end{equation}

In parallel with Section \ref{II}, we denote the solutions of the $\Lambda$-free FRGE as  $\EA{k}[\bar{h},\xi,\bar{\xi};\bar{g}]$.
According to equation \eqref{inicond0} for $k=\Lambda$, we get the corresponding relation:
\begin{equation}\label{qeginicond}
\EA{\Lambda,\Lambda}[\bar{h},\xi,\bar{\xi};\bar{g}]=\EA{\Lambda}[{h},\xi,\bar{\xi};\bar{g}]
\end{equation}

The integro-differential equation analogous to (\ref{integroeq}) reads in QEG:
\begin{align}\label{qeginteq}
\exp{\Big(-\EA{k,\Lambda}[\bar{h},\xi,\bar{\xi};\bar{g}]\Big)}= &
\int\mathcal{D}_{\Lambda}h\mathcal{D}_{\Lambda}C\mathcal{D}_{\Lambda}\bar{C}
\exp{\Big[
-\widetilde{S}_{\Lambda}[h,C,\bar{C};\bar{g}]}-
\nonumber\\ 
- & \Delta_k S[h-\bar{h},C-\xi,\bar{C}-\bar{\xi};\bar{g}] 
+\vol({h}_{\mu\nu}-\bar{h}_{\mu\nu})\frac{\delta\EA{k,\Lambda}}{\delta\bar{h}_{\mu\nu}}
\nonumber\\ 
+ & \vol(C^{\mu}-\xi^{\mu})\frac{\delta\EA{k,\Lambda}}{\delta\xi^{\mu}}
+\vol(\bar{C}_{\mu}-\bar{\xi}_{\mu})\frac{\delta\EA{k,\Lambda}}{\delta\bar{\xi}^{\mu}}\Big]
\end{align}

\subsection{The bare action at one loop}\label{oneloop}

As in the scalar case above, we would like to use the information contained in  a given solution
$\EA{k}[\bar{h},\xi,\bar{\xi};\bar{g}]$ of the $\Lambda$-free FRGE in order to find out which $\Lambda$-dependence must
be given to the (total) bare action $\widetilde{S}_{\Lambda}$ if we want the path integral to possess a well defined limit
$\Lambda\rightarrow\infty$ and to reproduce the prescribed $\EA{k}$. 
Using eq.\eqref{qeginicond} and \eqref{qeginteq} and restricting ourselves
to the 1-loop level, its derivation proceeds as in Section \ref{reconstructing}, with
the result
\begin{equation}\label{qegbareeff}
\EA{\Lambda,\Lambda}[\bar{h},\xi,\bar{\xi};\bar{g}]=\widetilde{S}_{\Lambda}[\bar{h},\xi,\bar{\xi};\bar{g}]
+\frac{1}{2}\textrm{STr}_{\Lambda}\ln{\Big[\Big(\widetilde{S}_{\Lambda}^{(2)}+\widehat{\mathcal{R}}_{\Lambda}\Big)[\bar{h},\xi,\bar{\xi};\bar{g}]\;\mathcal{N}^{-1}\Big]}
\end{equation}
Here $\mathcal{N}$ is a block diagonal normalization matrix, equal to $M^d$ and $M^2$ in the graviton and the ghost sector,
respectively.

\subsection{The twofold Einstein-Hilbert truncation}

Solving the above equation for the bare action $\widetilde{S}_{\Lambda}[\bar{h},\xi,\bar{\xi};\bar{g}]$ is  difficult,
even at the one-loop level, since \eqref{qegbareeff} is a complicated functional differential equation for the bare action.
In practice one has to restrict the space of actions where $\EA{k}$ and $\widetilde{S}_{\Lambda}$ are defined by 
truncating them  to a tractable number of terms. The simplest possibility, which we analyze here, is given by the 
Einstein-Hilbert truncation for both the effective and the bare action. As in \cite{mr}
we make the ansatz
%\subsection{The twofold Einstein-Hilbert truncation}\label{EHtruncation}
%
\begin{eqnarray}\label{EHansatz}
\EA{k}[g,\bar{g},\xi,\bar{\xi}] & = &-(16\pi G_k)^{-1}\vol\sqrt{g}\Big(R(g)-2\bar{\lambda}_k\Big) 
+S_{\textrm{gh}}[g-\bar{g},\xi,\bar{\xi};\bar{g}]
\nonumber\\
& & {} + (32\pi G_k)^{-1}\vol\sqrt{\bar{g}}\bar{g}^{\mu\nu}(\mathcal{F}_{\mu}^{\alpha\beta}g_{\alpha\beta})\!
(\mathcal{F}_{\nu}^{\rho\sigma}g_{\rho\sigma})
\end{eqnarray}
The third term on the RHS of eq.(\ref{EHansatz}) 
is the gauge fixing term\footnote{We employ a non-dynamical gauge fixing parameter $\alpha=1$ here.} corresponding to the harmonic coordinate
condition, involving $\mathcal{F}_{\mu}^{\alpha\beta}\equiv \delta^{\beta}_{\mu}\bar{g}^{\alpha\gamma}\bar{D}_{\gamma}
-\tfrac{1}{2}\bar{g}^{\alpha\beta}\bar{D}_{\mu}$, and the second term is the associated ghost action. 
We make an analogous ansatz for the bare action:
\begin{eqnarray}\label{EHbareansatz}
\widetilde{S}_{\Lambda}[g,\bar{g},\xi,\bar{\xi}] & = &-(16\pi \check{G}_{\Lambda})^{-1}\vol\sqrt{g}\Big(R(g)-2\check{\bar{\lambda}}_{\Lambda}\Big) 
+S_{\textrm{gh}}[g-\bar{g},\xi,\bar{\xi};\bar{g}]
\nonumber\\
& & {} + (32\pi \check{G}_{\Lambda})^{-1}\vol\sqrt{\bar{g}}\bar{g}^{\mu\nu}(\mathcal{F}_{\mu}^{\alpha\beta}g_{\alpha\beta})\!
(\mathcal{F}_{\nu}^{\rho\sigma}g_{\rho\sigma})
\end{eqnarray}
Eq.(\ref{EHansatz}) contains the running dimensionful parameters $G_k$ and $\bar{\lambda}_k$. The corresponding
bare Newton and cosmological constant, respectively, are denoted $\check{G}_{\Lambda}$ and $\check{\bar{\lambda}}_{\Lambda}$.

Setting the ghost terms to zero, $\xi=\bar{\xi}=0$, and $\bar{g}=g$,  the 
super trace has a derivative expansion of the form
\begin{equation}\label{qegDE}
\frac{1}{2}\textrm{STr}_{\Lambda}\ln{\Big[\Big(\widetilde{S}_{\Lambda}^{(2)}+\widehat{\mathcal{R}}_{\Lambda}\Big)[0,0,0;\bar{g}]\mathcal{N}^{-1}\Big]}\!=\!B_0\Lambda^d\vol\sqrt{g}+B_1\Lambda^{d-2}\vol\sqrt{g}R(g)+\cdots
\end{equation}
with dimensionless coefficients $B_0$ and $B_1$, respectively. In $d=4$ and using the optimized cutoff 
shape function they are given by \cite{elisa1}
\begin{subequations}\label{Bs}
\begin{align}
B_0 &= \frac{1}{32\pi^2}\Big[5\ln{(1-2\check{\lambda}_{\Lambda})}-5\ln{(\check{g}_{\Lambda})}  +Q_{\Lambda} \Big]
\label{b0}\\
B_1 &=\frac{1}{3}B_0+\Delta B_1
\label{b1}\\
\Delta B_1 &\equiv \frac{1}{16\pi^2}\frac{2-\check{\lambda}_{\Lambda}}{1-2\check{\lambda}_{\Lambda}}
\label{db}\\
Q_{\Lambda}&\equiv 12\ln{\big(\Lambda/M\big)}+b_0
\label{q}
\end{align}
\end{subequations}
with the constant $b_0\equiv -5\ln{(32\pi)}-\ln{2}$.
Using (\ref{qegDE}) in (\ref{qegbareeff}) and equating 
the coefficients of the independent invariants we obtain two equations relating the effective to the bare parameters:
\begin{align}\label{Eqcoup1}
\frac{1}{G_{\Lambda}}-\frac{1}{\check{G}_{\Lambda}} =  -16\pi\; B_1\;\Lambda^{d-2},\qquad &
\frac{\bar{\lambda}_{\Lambda}}{G_{\Lambda}}-\frac{\check{\bar{\lambda}}_{\Lambda}}{\check{G}_{\Lambda}}
= 8\pi\; B_0\;\Lambda^d
\end{align}
In terms of the dimensionless quantities defined by $g_{\Lambda}  \equiv  \Lambda^{d-2}G_{\Lambda}$, 
$\check{g}_{\Lambda}  \equiv  \Lambda^{d-2}\check{G}_{\Lambda}$, and analogous relations for the bare couplings, we get: 
\begin{subequations}\label{Eqcoup2}
\begin{align}
\frac{1}{g_{\Lambda}}-\frac{1}{\check{g}_{\Lambda}} & =  -16\pi \;B_1
\label{second}\\
\frac{{\lambda}_{\Lambda}}{g_{\Lambda}}-\frac{\check{\lambda}}{\check{g}_{\Lambda}}
& =  8\pi\; B_0
\label{third}
\end{align}
\end{subequations}
The algebraic system of equations \eqref{Eqcoup2} should allow  us to determine $\check{g}_{\Lambda}$ and $\check{\lambda}$ for
given $g_{\Lambda}$ and ${\lambda}_{\Lambda}$.

Unfortunately it is impossible to solve the system \eqref{Eqcoup2} analytically
for the bare parameters. However, in \cite{elisa1} we  solved those equations numerically, and found 
 a well defined pair
$(\check{g},\check{\lambda})$ for all $g>0$ and $\lambda<1/2$, for a wide range of values of 
the constant $Q=12\ln{c} +b_0$\footnote{The map  $(g,\lambda)\mapsto (\check{g},\check{\lambda})$
is explicitly $\Lambda$-dependent because of the parameter $ Q_{\Lambda}\equiv 12\ln{(\Lambda/M)}+b_0$.
This $\Lambda$-dependence can be removed by including appropriate factors of the
UV cutoff into the measure. If we set $M=c\Lambda$ with an arbitrary $c>0$ the quantity 
$Q=12\ln{c} +b_0$ becomes a $\Lambda$-independent constant. As a result, the map 
$(g,\lambda)\mapsto (\check{g},\check{\lambda})$ has no explicit dependence on any (UV or IR) cutoff.}.
Different values of $Q$ correspond to different normalizations of the measure.

Indeed, for an effective RG  trajectory, the fixed point behavior 
$\lim_{k\rightarrow\infty}(g_k,\lambda_k)=(g_*,\lambda_*)$ is mapped onto an analogous fixed 
point behavior at the bare level (after removing he explicit $\Lambda$ dependence from the map by setting $M=c\Lambda$).
The image of the GFP is always at $\check{g}_*=\check{\lambda}_*=0$,
while the coordinates of the ``bare'' NGFP, $\check{g}_*$ and $\check{\lambda}_*$, depend on the value
of $Q$. This behavior  is illustrated in  Fig.2, where we present the result of applying the map
$(g,\lambda)\mapsto (\check{g},\check{\lambda})$ for different values of $Q$, 
to a set of representative effective RG trajectories on the half plane $g>0$.
However, we emphasize  that all choices of $Q$ are 
physically equivalent. Varying Q simply amounts to shifting contributions back and forth
between the action and the measure.
\begin{figure}
\centering
\begin{tabular}{ll}
\bf{(a)} & \bf{(b)}\\
\includegraphics[width=5.5cm,height=3cm]{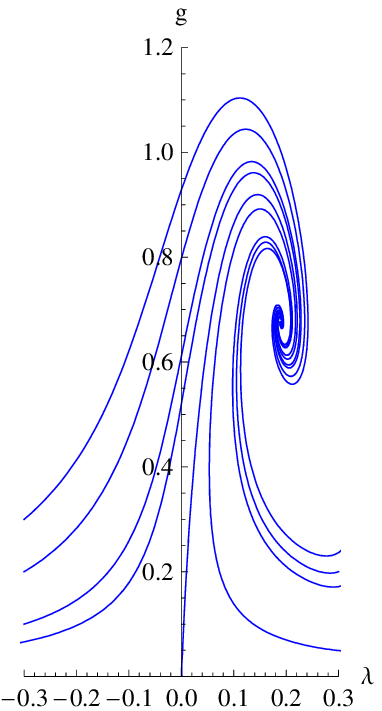} & \includegraphics[width=5.5cm,height=3cm]{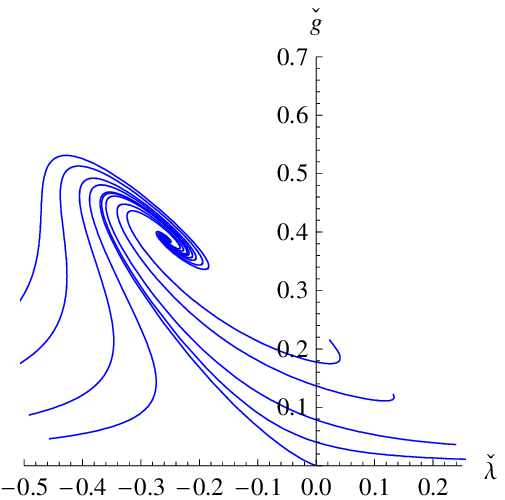}\\
 \bf{(c)}& \bf{(d)}\\
\includegraphics[width=5.5cm,height=3cm]{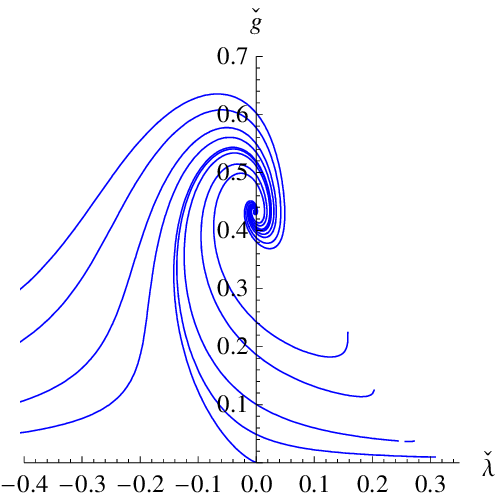} & \includegraphics[width=5.55cm,height=3cm]{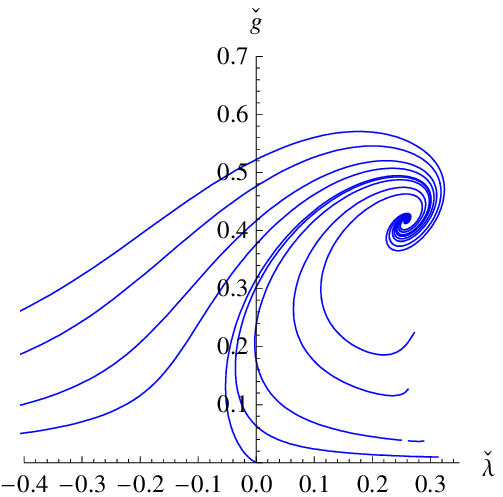}
\end{tabular}
\caption{\small{The diagram (a) shows the phase portrait of the effective RG flow on the $(g,\lambda)$-plane. The other diagrams are 
its image on the $(\check{g},\check{\lambda})$-plane of  bare parameters for three different values of $Q$, namely
(b) $Q=+1$ , (c) $Q=-0.1167$ where $\check{\lambda}_*=0$, and (d) $Q=-1$, respectively. }}
\end{figure}

It is instructive to determine the linearized flow near the two "bare" fixed points and to determine the corresponding
critical exponents, if they can be defined.

Both the ``effective'' and the ``bare''  NGFP are inner points of the corresponding
coupling constant space. The flow in the vicinity of one is the diffeomorphic image of the flow near the other. 
The RG running of the respective scaling fields is $\propto k^{-\theta}$ and 
$\propto \Lambda^{-\theta}$, respectively, with the same critical exponents $\theta$.
The ``bare'' GFP instead is located on  the line $\check{g}=0$, i.e. on the boundary  of the domain on which the map 
from the effective to the bare couplings
is defined. In its vicinity (on the half plane with $\check{g}>0$) the ``bare'' running is characterized by
logarithmically corrected power laws. The ``effective'' GFP, on the other hand, shows pure power
law scaling.
Near the GFP, we can expand the relations \eqref{Eqcoup2}, obtaining, in leading order:
\begin{align}\label{5.104}
 \check{g} =& g + \mathcal{O}({g}^2,{\lambda}^2)\\
 \check{\lambda} =& \lambda - \frac{{g}}{4\pi}\;\Big( Q-5\ln{{g}} \Big)
%\nonumber\\
+ \frac{{g}{\lambda}}{6\pi}\Big[ 3-Q+5\ln{{g}}\Big]
+ \mathcal{O}({g}^2,{\lambda}^2) 
\end{align}
These expansions are the first few terms of a power-log series. 
This implies that  the bare running indeed follows 
logarithmically corrected power laws.

\section{Conformally Reduced Gravity}\label{creh}

As another example of the strategy described above, 
we  next analyze conformally reduced gravity \cite{creh1,creh2} in which only the conformal factor 
of the metric is quantized. 
The simplicity of the model allows for the use of comparatively general truncations. 
We will use
the method of the Local Potential Approximation (LPA) to deduce the general form of the bare potential
contained in the reconstructed $\BA{\Lambda}$ of this model.

In conformally reduced gravity one considers only dynamical metrics $g_{\mu\nu}\equiv\phi^2\hat{g}_{\mu\nu}$ 
and background metrics $\bar{g}_{\mu\nu}\equiv\chi_B^2\hat{g}_{\mu\nu}$ 
which are conformal to a fixed reference metric $\hat{g}_{\mu\nu}$, usually taken to be $\hat{g}_{\mu\nu}=\delta_{\mu\nu}$.
The background metric is used in order to construct a coarse graining operator $\mathcal{R}_k[\chi_B]$ which cuts off the
spectrum of $-\bar{\square}$, the Laplace-Beltrami operator of $\bar{g}_{\mu\nu}$, at scale the $k^2$. In this way $1/k$ has
the character of a proper length with respect to the background metric, exactly as in full QEG. Furthermore, we introduce
a sharp UV cutoff by restricting the $-\bar{\square}$ eigenvalues to be smaller than $\Lambda^2$.
Following the same steps as in Section \ref{II}, one can construct a UV-regulated functional $W_{k,\Lambda}$ and with it
the corresponding effective average action $\EA{k,\Lambda}$. The reconstruction formula is a slight generalization of 
\eqref{finalequation}:
\begin{equation}\label{crehformula}
\EA{\Lambda,\Lambda}[\phi,\chi_B]-S_{\Lambda}[\phi,\chi_B]=\frac{1}{2}\rm{Tr}\Big[\theta(\Lambda^2+\bar{\square})
\ln{\Big\{ \big[ S^{(2)}_{\Lambda}[\phi,\chi_B]+\mathcal{R}_{\Lambda}[\chi_B] \big]M^{-2} \Big\} }\Big]
\end{equation}
Here, 
$S^{(2)}[\phi,\chi_B]_{xy}=\tfrac{1}{\sqrt{\hat{g(x)}}\sqrt{\hat{g(y)}}}\tfrac{\delta^2 }{\delta\phi(x)\delta\phi(y)}S[\phi,\chi_B]$,
and the explicit form of the coarse graining operator reads
\begin{equation}
\mathcal{R}_k[\chi_B]=-\frac{3}{4\pi G_k}\chi_B^2k^2R^{(0)}\Big( -\frac{\widehat{\square}}{\chi_B^2k^2}\Big)
\end{equation}

In order to solve \eqref{crehformula} we now make a  local potential  ansatz for both 
the effective and the bare action:
\begin{equation}\label{cinco}
\EA{k,\Lambda}[\phi,\chi_B]=-\frac{3}{4\pi G_{k,\Lambda}}\int\textrm{d}^4x\Big\{-\frac{1}{2}\phi\widehat{\square}\phi
+F_{k,\Lambda}(\phi,\chi_B)\Big\}
\end{equation}
\begin{equation}\label{seis}
\BA{\Lambda}[\phi,\chi_B]=-\frac{3}{4\pi \check{G}_{k,\Lambda}}\int\textrm{d}^4x\Big\{-\frac{1}{2}\phi\widehat{\square}\phi
+\check{F}_{k,\Lambda}(\phi,\chi_B)\Big\}
\end{equation}
Inserting \eqref{cinco} and \eqref{seis} into \eqref{crehformula} we get:
\begin{equation}\label{siete}
\frac{F_{\Lambda}(\phi,\chi_B)}{G_{\Lambda}}-\frac{\check{F_{\Lambda}}(\phi,\chi_B)}{\check{G_{\Lambda}}}=-\frac{1}{48\pi}\;\chi_B^4\;\Lambda^4\;\ln{
\Big[\frac{1}{\check{G_{\Lambda}}\Lambda^2}\big(\Lambda^2\chi_B^2+\partial^2_{\phi}\check{F}_{\Lambda}(\phi,\chi_B)\big)\Big]}
\end{equation}
We have derived the last equation {\it without setting } $\phi=\chi_B$, that is, $g_{\mu\nu}=\bar{g}_{\mu\nu}$. 
Our motivation is simply to keep 
separated  terms which are purely background dependent from those which are dynamical. 
The above truncations assume that the potentials have an extra, i.e. explicit dependence on $\chi_B$ 
(in addition to the one implicit in $\phi=\chi_B+\bar{f}$ where $\bar{f}$ is the fluctuation average).
Extended  truncations which have an  explicit dependence on the background, were 
 investigated  in this setting  in ref. \cite{elisa2}. 

It is convenient for the analysis to rewrite the above equation in terms of dimensionless quantities.
We use
\begin{align}\label{ocho}
g_{\Lambda}\equiv G_{\Lambda}\Lambda^2,&  \quad\varphi\equiv\Lambda\phi\quad\textrm{and}\quad b\equiv\Lambda\chi_B\\
Y_{\Lambda}(\varphi, b)\equiv& \Lambda^2F_{\Lambda}({\varphi}/{\Lambda},{b}/{\Lambda}) 
\end{align}
and analogous relations for the bare quantities. The resulting equation is:
\begin{equation}\label{nueve}
\frac{Y_{\Lambda}(\varphi, b)}{g_{\Lambda}}=\frac{\check{Y_{\Lambda}}(\varphi,b)}{\check{g_{\Lambda}}}-\frac{1}{48\pi}\;b^4\;\ln{
\Big[\frac{1}{\check{g_{\Lambda}}}\big(b^2+\partial^2_{\varphi}\check{Y}_{\Lambda}(\varphi,b)\big)\Big]}
\end{equation}
This equation strongly suggests that the bare potential $\check{Y_{\Lambda}}(\varphi,b)$ may depend explicitly on
the background field.

 Let us nonetheless start by exploring
 the ``$b=\varphi$" truncation which is analogous to the $g_{\mu\nu}=\bar{g}_{\mu\nu}$-truncation
used in QEG. Then eq.\eqref{nueve} reduces to 
\begin{equation}\label{diez}
\frac{Y_{\Lambda}(\varphi)}{g_{\Lambda}}=\frac{\check{Y_{\Lambda}}(\varphi)}{\check{g_{\Lambda}}}-\frac{1}{48\pi}\varphi^4\ln{
\Big[\frac{1}{\check{g_{\Lambda}}}\big(\varphi^2+\check{Y}^{\phantom{1}''}_{\Lambda}(\varphi)\big)\Big]}
\end{equation}
According to ref.\cite{creh2},  the $\Lambda$-free effective potential exhibits the following NGFP on the infinite 
dimensional space of the $Y$'s (for the ${R}^4$ topology):
\begin{subequations}
\begin{align}
Y_*(\varphi)= & -\frac{1}{6}\frac{\lambda_*}{g_*}\varphi^4\\
\lambda_*\approx 0.279, &\qquad g_* \approx 4.650
\end{align}
\end{subequations}
Therefore, one can insert this result on the LHS on eq.\eqref{diez} and solve for the bare potential $\check{Y_{*}}(\varphi)$.
It can be demonstrated that this indeed has a solution which can be found numerically.
 Asymptotically (for $\varphi\rightarrow\infty $) it  behaves as
\begin{equation}\label{asymptotic}
\check{Y_{*}}(\varphi)\approx \frac{\check{g}_*}{48\pi}\varphi^4\ln{\varphi^2} +\mathcal{O}(\varphi^4)
\end{equation}
Remarkably, while this potential is of the familiar Coleman-Weinberg form,
it is here part of the {\it{bare}} action; it  corresponds to a simple $\varphi^4$ monomial in the
effective one. Thus, as compared to a standard scalar matter field theory, the situation is exactly
inverted.

It is not difficult to understand how this comes about: The difference $\Gamma_*-S_*$ is given by a trace
$\textrm{Tr}[\cdots]$ which is nothing but a differentiated  one-loop determinant.
As a consequence, $\Gamma_*$ and $S_*$ differ precisely by terms typical of a one loop effective action,
and those include the potential term $\varphi^4\ln{ \varphi}$. Hence a $\varphi^4$ term in
$\Gamma_*$ unavoidably amounts  to a Coleman-Weinberg term in $S_*$, at least {\it within the
truncation considered}.
 
In fact, returning now to the more general truncations with an extra $\chi_B$-dependence 
of $F_k(\phi,\chi_B)$ it can be shown that actually $S_*$ and $\Gamma_*$ do not differ by a ``dynamical"
term $\varphi^4\ln{\varphi}$, nonanalytic in the quantum field, but rather merely by its background analog $b^4\ln{b}$.
It also can be shown \cite{elisa2} that the bare potential is analytic in $\varphi$ if the effective one is so.
This example nicely demonstrates that occasionally the oversimplifications caused by the class of ``$b=\varphi$", or
``$g_{\mu\nu}=\bar{g}_{\mu\nu}$" truncations can lead to a qualitatively wrong picture.

%%%%%
%%%%%

\section{Discussion}\label{sectionsix}

Here we described some first steps towards solving the reconstruction
problem for asymptotically safe quantum field theories. In particular
we showed explicitly that, after specifying a UV regularization
scheme and a measure, every solution of the flow equation for the effective average action 
without an UV cutoff gives rise to a regularized path integral with
a well defined limit $\Lambda\rightarrow\infty$, and to a UV cutoff dependent
bare action.

While the method we developed is completely general, this
work was motivated by the Asymptotic Safety program in Quantum Einstein Gravity. As to yet the 
investigations based upon the EAA focused on computing RG trajectories of the $\Lambda$-free FRGE and
establishing the existence of a non-Gaussian fixed point. The present work aims at completing the 
Asymptotic Safety program in the sense of finding the, yet unknown, quantum system which we
implicitly quantize by picking a solution of the flow equation. In fact, in our approach the primary
definition of ``QEG'' is in terms of an RG trajectory of the EAA that emanates from the fixed point.
The advantage of this strategy, defining the theory in terms of an effective rather than bare action,
is that it automatically guarantees an ``asymptotically safe'' high energy behavior. The
disadvantage is that in order to complete the Asymptotic Safety program, that is, to find the 
underlying microscopic theory, extra work is needed.

Once we know the microscopic, i.e. bare action we can 
attempt a kind of ``Legendre transformation'' to find appropriate phase space variables, a
microscopic Hamiltonian, and thus a canonical description of the bare theory. Only 
at this level we can identify the degrees of freedom that got quantized, as well as their fundamental 
interactions. Since the Hamiltonian is unlikely to turn out quadratic in the momenta, the 
``Legendre tansformation'' involved is to be understood as a generalized, i.e. quantum
mechanical one. In the simplest case it consists in reformulating a given configuration space path integral 
$\int\mathcal{D}\,\Phi \exp{\big(iS[\Phi]\big)}$ as a phase space integral
$\int\mathcal{D}\Phi\int\mathcal{D}\Pi \exp{\big(i\int\Pi\dot{\Phi}-H[\Pi,\Phi]\big)}$.
With other words, we must undo the integrating out of the momenta.

However, given the complexity of $\Gamma_*$ which most probably contains higher derivatives and non-local
terms a generalized, Ostrogradski-type phase space formalism will emerge presumably.

Being interested in a canonical description of the ``bare'' NGFP action one might wonder if there
exists an alternative formalism which deals directly with the RG flow of Hamiltonians rather
than Lagrangians.
It seems that there hardly can be a practicable approach of this kind which is  similar in spirit to the
EAA. The reason is as follows. 

If we apply a coarse graining step to an action which contains only, say, 
first derivatives of the field, the result will contain higher derivatives in general. This poses
no special problem in a Lagrangian setting, but for the Hamiltonian formalism it implies that
new momentum variables must be introduced. As a result, the coarse grained Hamiltonian
``lives'' on a different phase space (in the sense of Ostrogradski's method) than the original one.
Therefore, at least in a straightforward interpretation, there is no Hamiltonian analog of the 
flow on the space of actions. For this reason there is probably no simple way of getting around
the ``reconstruction problem''.

However, the above discussion does not contradict other approaches where the renormalization 
procedure could be applied
in a Hamiltonian description \cite{zapcor} since there the coarse graining is  performed in space 
(rather than spacetime) only. 

One should also emphasize that it is by no means clear from the 
outset what kind of fundamental degrees of freedom will be found in this Hamiltonian analysis. 
In our approach {\it the only nontrivial input is the theory space}, the space of functionals on which
the renormalization group operates. Having fixed this space a FRGE can be written down, 
the resulting flow can be computed, its fixed point(s) $\Gamma_*$  can be identified, and the associated asymptotically safe
field theories can  be defined {\it{without any additional input}}. As a consequence, the only statement about
the degrees of freedom in these theories which we can make on general grounds is that they can be 
``carried'' by precisely those fields on which $\Gamma_k$ depends.
(In the case at hand, theory space contains all functionals $\Gamma[g,\bar{g},\xi,\bar{\xi}]$ which 
are invariant under diffeomorphisms.) Clearly, just knowing the carrier field but not the action, here 
$\Gamma_*$, tells us comparatively little about the degrees of freedom. The action $\Gamma_*$, however, 
is a {\it{prediction}} of the theory,
not an input. From this point of view it is quite nontrivial that QEG was found to have  RG trajectories which
indeed describe classical General Relativity on macroscopic scales.

In this work we also investigated  QEG in the Einstein-Hilbert truncation, 
constructing a map relating the effective to the bare Newton  and cosmological constant, and we
analyzed the properties of the ``bare'' RG flow. We saw in particular that the ``effective'' NGFP
maps onto a corresponding ``bare'' one; in its vicinity the scaling fields show a power law running with the
same critical exponents as at the effective level. The situation is different for the GFP which is a 
boundary point of parameter space. The pure power laws of the ``effective'' flow receive logarithmic
corrections on the ``bare'' side. We also described the case of conformally reduced gravity 
within an (infine dimensional!) truncation of the LPA type. In this example we saw in particular that in 
order to get a qualitatively correct picture one must go beyond the class of ``$g=\bar{g}$"-truncations.

Leaving aside gravity,
in  future work it will be interesting to analyze for instance also higher dimensional Yang-Mills theory along the same 
lines. In fact, in ref.\cite{nonabavact} the effective average action of $d$-dimensional Yang-Mills theory
was considered in a simple $\int (F^a_{\mu\nu})^2$-truncation. According to this truncation\footnote{
For a generalization see also \cite{giesymfp}.}, $\Gamma_k$ has a NGFP in the UV if $4<d<24$.
Inspired by the structure of the one-loop effective action in Yang-Mills  theory one would expect that the 
``bare'' counterpart of the $\int (F^a_{\mu\nu})^2$-fixed point should contain terms like 
$\int (F^a_{\mu\nu})^2\ln{(F^a_{\mu\nu})^2}$, and also nonlocal ones such as 
$\int F^a_{\mu\nu}f(-D^2)F^a_{\mu\nu}$. For the following reason it is of some practical 
importance to find out whether this is actually the case in a sufficiently general, reliable truncation.
It seems comparatively easy to perform Monte-Carlo simulations in $d=5$, say, so that one could
possibly get an independent confirmation of the results obtained from the average action.
However, the problem is that a priori we do not really know which bare theory should be simulated
in order to arrive at the lattice version of the average action results. The present analysis suggests
 that if  Yang-Mills theory is asymptotically safe in $d=5$, the effective  fixed point action
$\Gamma_*$ might be simple, but $S_*$ could contain ``exotic'' nonlinear and nonlocal terms. If so, it
is conceivable that $S_*$ is sufficiently different from $\int (F^a_{\mu\nu})^2$ to belong to a new
universality class. In this case a Monte-Carlo simulation based upon the conventional Wilson gauge
field action might not find a NGFP, while it should show up when a discretized version of $S_*$ is used.

Completely analogous remarks apply to the nonlinear sigma model in $d>2$ which, according to the lowest
order truncation of the EAA, is asymptotically safe too \cite{nonlinsig}.

%\aknowledgments

\appendix

\section{The induced cosmological constant, and \\
what we can  learn from it}\label{COSMO}

In this appendix we illustrate  how the bare and the average action are related by means of a simple example: the cosmological constant 
induced by a scalar matter field quantized in a classical gravitational background. The example also serves as a toy model to highlight
several issues arising in the complete formulation of QEG. For further details we refer to \cite{elisa1}

We start with an action of a scalar field which is minimally coupled with the classical metric $g_{\mu\nu}$.
As we are interested only in the induced cosmological constant it will be sufficient to
 keep the $\vol\sqrt{g}$ gravitational invariant  in the bare and average 
action, respectively:
\begin{equation}\label{cosmobareaction}
S_{\Lambda}[\chi]=\frac{1}{2}\vol\sqrt{g}\Big[g^{\mu\nu}\partial_{\mu}\chi\partial_{\nu}\chi +\check{m}^2\chi^2\Big] +
\check{C}_{\Lambda}\vol\sqrt{g}
\end{equation}
\begin{equation}\label{cosmoeffaction}
\EA{k,\Lambda}[\phi]=\frac{1}{2}\vol\sqrt{g}\Big[g^{\mu\nu}\partial_{\mu}\phi\partial_{\nu}\phi +{m}^2\phi^2\Big] +
{C}_{k,\Lambda}\vol\sqrt{g}
\end{equation}
The solution $\EA{k}[\phi]$ of the $\Lambda$-free FRGE has a structure similar to (\ref{cosmoeffaction}) involving a
running parameter $C_k$. The three $C$-factors $\check{C}_{\Lambda}$, $C_{k,\Lambda}$ and $C_k$ are related to the corresponding
cosmological constants $\bar{\lambda}$ by $C\equiv (\bar{\lambda}/8\pi G)$
where Newton's constant $G$ does not run in the approximation considered. Furthermore, for the purposes of this demonstration, the running of the masses is also neglected.

Notice that since $S_{\Lambda}$ is quadratic in $\chi$ the functional integral (\ref{wkl}) for $W_{k,\Lambda}[J]$, appropriately 
generalized to a curved background, can be solved exactly.
 With the restricted trace $\rm{Tr}_{\Lambda}[\cdots]\equiv \rm{Tr}[\theta(\Lambda^2+D^2)(\cdots)]$
one obtains
\begin{eqnarray}
W_{k,\Lambda}[J] &=& \frac{1}{2}\vol\sqrt{g}\; J\; \big[  -D^2+ \check{m}^2+\mathcal{R}_k(-D^2)\big]^{-1} J-
\check{C}_{\Lambda}\vol\sqrt{g}-
\nonumber\\
& & {}-\frac{1}{2}\textrm{Tr}_{\Lambda}\ln{\Big[\Big( -D^2+ \check{m}^2+\mathcal{R}_k(-D^2)\Big)M^{-2}\Big]}
\end{eqnarray}
In this simple case we can compute $\EA{k,\Lambda}$ directly from the very definition of the EAA, eq.(\ref{effectiveaction}):
\begin{eqnarray}\label{cosmoeffaction2}
\EA{k,\Lambda}[\phi] &=& \frac{1}{2}\vol\sqrt{g}\Big(g^{\mu\nu}\partial_{\mu}\phi\partial_{\nu}\phi +\check{m}^2\phi^2 \Big)+
\check{C}_{\Lambda}\vol\sqrt{g}+
\nonumber\\
& & {} + \frac{1}{2}\textrm{Tr}_{\Lambda}\ln{\Big[\Big( -D^2+ \check{m}^2+\mathcal{R}_k(-D^2)\Big)M^{-2}\Big]}
\end{eqnarray}

The flow equation for $\EA{k,\Lambda}[\phi]$ is a slight generalization of (\ref{frge}) with the flat metric replaced by 
$g_{\mu\nu}$ everywhere. In particular, the operator $\hat{p}^2\equiv -D^2$ is now
to be interpreted as the Laplace-Beltrami operator constructed
with the metric $g_{\mu\nu}$. Upon inserting (\ref{cosmoeffaction}) the  FRGE assumes the form
\begin{equation}\label{cosmofrge}
k\partial_kC_{k,\Lambda}\vol\sqrt{g}=\frac{1}{2}\textrm{Tr}\Big[\theta(\Lambda^2+D^2)\;\mathcal{K}(-D^2)^{-1}\;
k\partial_k\mathcal{R}_k(-D^2)\Big]
\end{equation}
with $ \mathcal{K}(\hat{p}^2)\equiv \hat{p}^2+m^2+\mathcal{R}_k(\hat{p}^2)$.
To make eq.(\ref{cosmofrge}) consistent we may retain only the volume term $\propto$  $\vol\sqrt{g}$ in the derivative 
expansion of the trace on its RHS. It is easily found by inserting a flat metric. Using the optimized cutoff  
\eqref{cosmofrge} it  reduces to, with $v_d\equiv [2^{d+1}\pi^{d/2}\Gamma(d/2)]^{-1}$,
\begin{equation}\label{cosmofrge1}
k\partial_kC_{k,\Lambda}=\frac{4v_d}{d}\Big(\frac{k^2}{k^2+m^2}\Big)\;k^d
\end{equation}
We observe that the RHS of (\ref{cosmofrge1}) has become {\it independent of the cutoff $\Lambda$}. 

Inserting the 
$\EA{k}$-ansatz (involving $C_k$) into the $\Lambda$-free flow equation we find  eq.(\ref{cosmofrge1}), too,
 this time for $C_k$. 
Hence $k\partial_kC_k=k\partial_kC_{k,\Lambda}$ for all $ k\leq \Lambda$, and therefore $C_k=C_{k,\Lambda}$
for $k\leq \Lambda$ if the same initial conditions are imposed on $C_k$ and $C_{k,\Lambda}$.

If $k\gg m$, eq.(\ref{cosmofrge1}) yields the familiar $k^d$-running of the cosmological constant; it is this scale 
dependence that would  result from summing up the zero point energies of the (massless) field modes. If $k\ll m$
 the running is much weaker since the RHS of (\ref{cosmofrge1}) contains a suppression factor $(k/m)^2\ll 1$. 
This is a typical decoupling phenomenon: In the regime $k\ll m$ the physical mass $m$ is the active IR cutoff.

The RG equation (\ref{cosmofrge1}) has the solution
\begin{equation}\label{cosmosol}
C_{k,\Lambda}= C_{\textrm{ren}}+\frac{2v_d}{d}\int^{k^2}_{0}\!\!\textrm{d}y\frac{y^{\frac{d}{2}}}{y+m^2}
\end{equation}
Here we fixed a specific RG trajectory by imposing the renormalization condition 
\protect{$C_{k=0,\Lambda\rightarrow\infty}=C_{\textrm{ren}}$} with $\bar{\lambda}_{\textrm{ren}}\equiv (8\pi G)C_{\textrm{ren}}$ the 
``renormalized cosmological constant", to be determined experimentally in principle. For $m=0$ in particular, since 
$C_k=C_{k,\Lambda}$ for $k$ below $\Lambda$,
\begin{equation}\label{cosmosolA}
C_k=C_{k,\Lambda}=C_{\textrm{ren}}+ 4d^{-2}\;v_d\; k^d
\end{equation}
If $d=4$, say, in standard notation,
\begin{equation}\label{cosmosolB}
\bar{\lambda}_{k}=\bar{\lambda}_{\textrm{ren}}+\frac{1}{16\pi^2}\;G_0\; k^4
\end{equation}
The scalar being massless, this running of the effective cosmological constant has the same structure as in
pure quantum gravity \cite{mr}.

By performing a derivative expansion of $\textrm{Tr}_{\Lambda}\ln{[\cdots]}$ in (\ref{cosmoeffaction2}) we can obtain the
scalar's contribution to the induced cosmological constant ($\int\sqrt{g}$ term), the induced Newton constant ($\int\sqrt{g}R$
term), and similarly to the higher derivative terms. Here we are interested in the cosmological constant only, and comparing
(\ref{cosmoeffaction2}) to (\ref{cosmoeffaction}) yields
\begin{eqnarray}\label{cosmodiff}
C_{k,\Lambda}-\check{C}_{\Lambda} & = &\frac{1}{2} \Big[\vol\sqrt{g}\Big]^{-1}\;\textrm{Tr}_{\Lambda}\ln{\Big[\cdots\Big]}\Big|_{\int\sqrt{g} \textrm{ term}}
\nonumber\\
& & {} = \frac{1}{2}\int\!\!\frac{\textrm{d}^dp}{(2\pi)^d}\;\theta(\Lambda^2-p^2)\;\ln{\Big(\big[p^2+m^2 +\mathcal{R}_k(p^2)\big]M^{-2}\Big)}
\end{eqnarray}
Employing the optimized cutoff again, (\ref{cosmodiff}) evaluates to
\begin{equation}\label{cosmodiff2}
C_{k,\Lambda}=\check{C}_{\Lambda} +\frac{2v_d}{d}\; k^d\;\ln{\Big(\frac{k^2+m^2}{M^2}\Big)}
+v_d\int_{k^2}^{\Lambda^2}\!\!\textrm{d}y \;y^{d/2-1}\ln{\Big(\frac{y^2+m^2}{M^2}\Big)}
\end{equation}
Note that in (\ref{cosmodiff}) and (\ref{cosmodiff2}) we replaced $\check{m}$ with $m$ since comparing the $\phi^2$-terms in 
(\ref{cosmoeffaction2}) and (\ref{cosmoeffaction}), respectively, implies that $\check{m}=m$ within the simple truncation used. 

For $m=0$ and $d=4$, say, eq.(\ref{cosmodiff2}) implies the following explicit result for the running effective cosmological
constant in terms of the bare one:
\begin{equation}
C_{k,\Lambda}=\check{C}_{\Lambda}+v_4\Big[  \Lambda^4\ln{(\Lambda/M)}-\frac{1}{4}(\Lambda^4-k^4)\Big]
\end{equation}
For arbitrary $d$ and $m$, the limit $k\rightarrow\Lambda$ of eq.(\ref{cosmodiff2}) reads
\begin{equation}\label{barecc0}
\check{C}_{\Lambda}=C_{\Lambda,\Lambda}-\frac{2v_d}{d}\;\Lambda^d\;\ln{\Big(\frac{\Lambda^2+m^2}{M^2}\Big)}
\end{equation}
This equation tells us how, for a given effective cosmological constant $C_{\Lambda,\Lambda}$, the bare one, 
$\check{C}_{\Lambda}$, must be adjusted in order to give rise to the prescribed effective one. The value
of $C_{\Lambda,\Lambda}$ in turn depends on the RG trajectory chosen, i.e., in this simple situation, on the value of 
$C_{\textrm{ren}}$. In fact, from the explicit solution (\ref{cosmosol}) we get
\begin{equation}\label{effcc0}
C_{\Lambda,\Lambda}= C_{\textrm{ren}}+\frac{2v_d}{d}\int^{\Lambda^2}_{0}\!\!\textrm{d}y\frac{y^{\frac{d}{2}}}{y+m^2}
\end{equation}

The above simple formulae illustrate various conceptual lessons of general significance.
The first lesson we illustrate with this model is the non-uniqueness of the bare action.
For the massless case $m=0$ in eq.\eqref{barecc0}, the cosmological constant in the bare action is
\begin{equation}\label{barecc}
\check{C}_{\Lambda}=C_{\Lambda,\Lambda}-4d^{-1}v_d\;\Lambda^d\;\ln{(\Lambda/M)}
\end{equation}
while the one in $\EA{k,\Lambda}$ and $\EA{k}$ at $k=\Lambda$ reads
\begin{equation}\label{effcc}
C_{\Lambda,\Lambda}= C_{\textrm{ren}}+ 4d^{-2}v_d\;\Lambda^d=C_{k=\Lambda}
\end{equation}
It is clear from here that choosing different values of the free parameter  $M$  
 will affect the bare cosmological constant (\ref{barecc}) but not the effective one, eq.(\ref{effcc}).
 The {\it{effective}} cosmological constant $C_{k=\Lambda}$ will always be
proportional to $\Lambda^d$ for $\Lambda\rightarrow\infty$ and approach {\it{plus}} infinity.

As a first choice consider $M=\textrm{const}$, i.e. $M$ is a positive constant independent of $\Lambda$. Then, according to
(\ref{barecc}), the {\it{bare}} cosmological constant $\check{C}_{\Lambda}$ is proportional to $-\Lambda^d\ln{\Lambda}$ for
$\Lambda\gg M$ and it approaches {\it{minus}} infinity in the limit $\Lambda\rightarrow\infty$.

As a second choice assume $M$ is proportional to the UV cutoff, $M=c\Lambda$, with some constant $c>0$. Then 
$\check{C}_{\Lambda}=C_{\textrm{ren}}+ 4d^{-2}v_d\Lambda^d\{1-d\ln{c}\}$ diverges proportional to $\Lambda^d$ if 
$c\neq\exp{(1/d)}$, and depending on the value of $c$ it might approach  $-\infty$ or $+\infty$.
In the special case $c=\exp{(1/d)}$ the bare cosmological constant $\check{C}_{\Lambda} $ equals $C_{\textrm{ren}}$ for all
$\Lambda$, i.e. it is {\it{finite}} even in the limit $\Lambda\rightarrow\infty$. Also $c=1$ is special: in this case,
accidentally, the bare and the effective average action contain {\it{the same}} cosmological constant: 
$\check{C}_{\Lambda}=C_{\Lambda,\Lambda}$.

Even though they can lead to dramatically different bare actions, the various choices for $M$ are all physically 
equivalent.
The ordinary effective action  and the EAA are independent of $M$. Changing $M$ simply
amounts to shifting contributions from the measure into the bare action or vice versa.

This illustrates a general lesson which, while true everywhere in quantum field theory, is particularly important in the 
asymptotic safety context: It makes no sense to talk about a bare action unless one has specified a measure before;
neither $\mathcal{D}_{\Lambda}\chi$ nor $\exp{[-S_{\Lambda}]}$ have a physical meaning separately, only the combination
$\int\mathcal{D}_{\Lambda}\chi\exp{[-S_{\Lambda}]}$ has. Here we illustrated this phenomenon by a simple rescaling
of the integration variable but clearly it extends to more general transformations of $\chi$ whose Jacobian is interpreted
as changing the action $S_{\Lambda}$ to a new one, $S^{'}_{\Lambda}$. 

The concrete lesson for the asymptotic safety program is that one should not expect a fixed point solution of the FRGE,
$\Gamma_*$, to correspond to a unique bare action.

Also, a natural question to ask  is if  there is a flow  equation that governs the $\Lambda$-dependence of the bare 
actions defined with our strategy. For the present toy model, this flow equation can be easily derived using \eqref{barecc0}: 
\begin{equation}\label{cosmoRGeq}
\Lambda\partial_{\Lambda}\check{C}_{\Lambda}=-\frac{4v_d}{d}\;\Lambda^d\;\Big[\frac{d}{2}\ln{\Big(\frac{\Lambda^2+m^2}{M^2}\Big)}     
-\frac{\Lambda\partial_{\Lambda}M}{M}\Big]
\end{equation}
This equation tells how the bare action must change when $\Lambda$ is sent to infinity, given the requirement that the
parameter $C_{k=0}$ in the ordinary effective action assumes the prescribed value $C_{\textrm{ren}}$. 

Obviously
{\it{the RG equation for the bare cosmological constant
 is quite different from the corresponding equation at the level of the effective average action}},
eq.(\ref{cosmofrge1}). 

So, for constructing a path integral describing an asymptotically safe theory, why not use a full fledged functional flow
equation for the bare action? Why is the RG flow of $\EA{k}$  crucial for the QEG program, while $S_{\Lambda}$ plays only
a secondary role? There are at least two answers to these questions:

The first answer is that {\it{the property of asymptotic safety 
is decided about at the
effective rather than bare level.}} By its very definition, asymptotic safety requires observable quantities
such as scattering cross sections to be free from  divergences. Since the $S$-matrix elements are 
essentially functional derivatives of $\Gamma\equiv \Gamma_{k=0}$ this requires the ordinary effective
action to be free from such divergences. This is indeed the case if $\Gamma$ is connected to a UV fixed point 
$\Gamma_*$ by a regular RG trajectory. So, in order to test wether this condition is satisfied we need to 
know the $\Gamma_k$-flow. The concomitant $S_{\Lambda}$-flow is of no direct physical relevance.
In principle is is even conceivable that, while $\Gamma_k$ approaches to a fixed point  in the UV, the bare action
does not; the resulting theory could nevertheless have completely acceptable physical properties.

For these reasons the basic tool in searching for asymptotic safety is the flow equation for the EAA
and not its analog for the bare action.

A second answer to the above question is that we
 would like the scale dependent functional obtained by solving the flow equation 
to have a chance of defining an effective field theory in the sense that its tree level evaluation at some scale 
approximately describes all quantum effects with this typical scale. For $\EA{k}$ this is indeed the 
case\footnote{Of course we are not saying here that $\EA{k}$ necessarily provides a numerically precise description. To what degree
this is actually possible (fluctuations are small, etc.) depends on the details of the physical situation.}, 
but not for $S_{\Lambda}$. The reason is that, given $S_{\Lambda}$, there is still a functional integration to be performed in 
order to go over to the effective level; using $\EA{k}$ instead, it has been performed already.

The above toy model illustrates this point: From eq.(\ref{cosmosolA}) or eq.(\ref{cosmosolB}) we conclude that for every finite
$\bar{\lambda}_{\textrm{ren}}\equiv (8\pi G)C_{\textrm{ren}}$ the {\it{running effective}} cosmological constant
$\bar{\lambda}_k\equiv (8\pi G)C_{k}$ becomes large and positive for growing $k$ and finally approaches {\it{plus}} infinity
for $k\rightarrow\infty$. Applying the effective field theory interpretation we would  insert this $\bar{\lambda}_k$ into the effective Einstein equation. It then 
predicts that, at high momentum scales, spacetime  is strongly curved and has {\it{positive}} curvature. 

From the above remarks  it is clear that the {\it{running bare}} action does not contain this information.
Depending on our choice for $M$ the bare cosmological constant $\check{C}_{\Lambda}$ approaches to $+\infty$,$-\infty$ or a finite value where $\Lambda\rightarrow\infty$. So clearly it would not make any sense to insert it
into Einstein's equation in order to ``RG improve" it.


\begin{thebibliography}{99} 
\bibitem{kiefer}
For  general introductions see C.~Kiefer, \emph{Quantum Gravity}, Second Edition, \\
Oxford Science Publications, Oxford (2007);
H.~Hamber, \emph{Quantum Gravitation}, Springer, Berlin (2008).
%
\bibitem{A}
A.~Ashtekar, \emph{Lectures on non-perturbative canonical gravity},\\
World Scientific, Singapore 1991; \\
A.~Ashtekar and J.~Lewandowski, \emph{Background independent quantum gravity: A status report,},
\emph{Class.\ Quant.\ Grav.} {\bf{21}}  (2004) R53 [{\tt gr-qc/0404018}].
%
%
\bibitem{R}
C.~Rovelli, \emph{Quantum Gravity}, Cambridge University Press, Cambridge 2004.
%
\bibitem{T}
Th.~Thiemann, \emph{Modern Canonical Quantum General Relativity},\\
Cambridge University Press, Cambridge 2007.
%
\bibitem{wein}
S.~Weinberg 
in \emph{General Relativity, an Einstein Centenary Survey},\\
S.W.~Hawking and W.~Israel (Eds.), 
Cambridge University Press 1979;\\
S.~Weinberg,\emph{What is quantum field theory, and what did we think it was?},
\mbox{[{\tt hep-th/9702027}]}; 
 \emph{Living with Infinities}, \mbox{[\tt 0903.0568 [hep-th]]}. 
 

%
\bibitem{mr}
M.~Reuter, \emph{Nonperturbative Evolution Equation for Quantum Gravity},
Phys.\ Rev.\ D {\bf 57} (1998) 971, \mbox{[{\tt hep-th/9605030}]}.
%
\bibitem{percadou}
D.~Dou and R.~Percacci, \emph{The running gravitational couplings},
Class.\ Quant.\ Grav. {\bf{15}} (1998) 3449, \mbox{[{\tt hep-th/9707239}]}.
%
\bibitem{oliver1}
O.~Lauscher and M.~Reuter, \emph{Ultraviolet fixed point and generalized flow equation of quantum
  gravity},
  Phys.\ Rev.\  D {\bf 65} (2002) 025013
  \mbox{[{\tt hep-th/0108040}]}.
%
\bibitem{frank1}
M.~Reuter and F.~Saueressig, \emph{Renormalization group flow of quantum gravity in the Einstein-Hilbert
  truncation},
  Phys.\ Rev.\  D {\bf 65} (2002) 065016
  \mbox{[{\tt hep-th/0110054}]}.
%
\bibitem{oliver2}
O.~Lauscher and M.~Reuter, \emph{Flow equation of quantum Einstein gravity in a higher-derivative
  truncation},
  Phys.\ Rev.\  D {\bf 66}, (2002) 025026 
\mbox{[{\tt hep-th/0205062}]}.
%
\bibitem{oliver3}
O.~Lauscher and M.~Reuter, \emph{Is quantum Einstein gravity nonperturbatively renormalizable?},
  Class.\ Quant.\ Grav.\  {\bf 19}, (2002) 483 
\mbox{[{\tt hep-th/0110021}]}.
%
\bibitem{oliver4}
O.~Lauscher and M.~Reuter, \emph{Towards nonperturbative renormalizability of quantum Einstein gravity},
  Int.\ J.\ Mod.\ Phys.\  A {\bf 17},(2002) 993 
\mbox{[{\tt hep-th/0112089 }]}.
%
\bibitem{souma}
W.~Souma, \emph{Non-trivial ultraviolet fixed point in quantum gravity},
  Prog.\ Theor.\ Phys.\  {\bf 102},(1999) 181 
 \mbox{[{\tt{hep-th/9907027}}]}.
%
\bibitem{frank2}
M.~Reuter and F.~Saueressig, \emph{A class of nonlocal truncations in quantum Einstein gravity and its
  renormalization group behavior},
  Phys.\ Rev.\  D {\bf 66},(2002)  125001 \mbox{[{\tt{hep-th/0206145}}]};\\
 M.~Reuter and F.~Saueressig,
 \emph{Nonlocal quantum gravity and the size of the universe},
  Fortsch.\ Phys.\  {\bf 52} (2004) 650
 \mbox{[{\tt{hep-th/0311056}}]}.
%
\bibitem{prop}
A.~Bonanno and M.~Reuter,
  \emph{Proper time flow equation for gravity},
  JHEP {\bf 0502},(2005) 035 
\mbox{[{\tt hep-th/0410191}]}.
%
\bibitem{oliverbook}
For reviews see: 
M.~Reuter and F.~Saueressig, \emph{Functional Renormalization Group Equations, Asymptotic Safety, and Quantum
  Einstein Gravity},
\mbox{[{\tt 0708.1317 [hep-th]}]};\\
O.~Lauscher and M.~Reuter in \emph{Quantum Gravity}, B.~Fauser, \\
J.~Tolksdorf and E.~Zeidler (Eds.), Birkh\"auser, Basel 2007  \mbox{[{\tt hep-th/0511260 }]};\\
O.~Lauscher and M.~Reuter in \emph{Approaches to Fundamental Physics}, \\
I.-O.~Stamatescu and E.~Seiler (Eds.), Springer, Berlin 2007.

%
\bibitem{perper1}
R.~Percacci and D.~Perini,
  \emph{Constraints on matter from asymptotic safety},
  Phys.\ Rev.\  D {\bf 67} (2003) 081503 \mbox{[{\tt{hep-th/0207033}}]};
\emph{Asymptotic safety of gravity coupled to matter},
  Phys.\ Rev.\  D {\bf 68} (2003) 044018 \mbox{[{\tt{hep-th/0304222}}]};
\emph{Should we expect a fixed point for Newton's constant?},
  Class.\ Quant.\ Grav.\  {\bf 21},(2004) 5035 \mbox{[{\tt{hep-th/0401071}}]}.
%
\bibitem{codello}
A.~Codello and R.~Percacci, \emph{Fixed Points of Higher Derivative Gravity},
  Phys.\ Rev.\ Lett.\  {\bf 97} (2006) 221301\mbox{[{\tt{hep-th/0607128}}]};\\
A.~Codello, R.~Percacci and C.~Rahmede, \emph{Ultraviolet properties of f(R)-gravity},
  Int.\ J.\ Mod.\ Phys.\  A {\bf 23} (2008) 143.

%
\bibitem{litimgrav}
D.~Litim, \emph{Fixed points of quantum gravity},
  Phys.\ Rev.\ Lett.\  {\bf 92} (2004) 201301
  \mbox{[{\tt{hep-th/0312114}}]};
\emph{On fixed points of quantum gravity},
  AIP Conf.\ Proc.\  {\bf 841},(2006) 322 
\mbox{[{\tt{hep-th/0606044}}]};\\
P.~Fischer and D.~Litim, \emph{Fixed points of quantum gravity in extra dimensions},
  Phys.\ Lett.\  B {\bf 638}, (2006) 497 
\mbox{[{\tt{hep-th/0602203}}]};
\emph{Fixed points of quantum gravity in higher dimensions},
  AIP Conf.\ Proc.\  {\bf 861},(2006) 336
\mbox{[{\tt{hep-th/0606135}}]}.
%
\bibitem{frankmach}
P.~Machado and F.~Saueressig, \emph{On the renormalization group flow of f(R)-gravity},
  Phys.\ Rev.\  D {\bf 77},(2008) 124045 
  \mbox{[{\tt{0712.0445 [hep-th]}}]}.

%
\bibitem{oliverfrac}
O.~Lauscher and M.~Reuter, \emph{Fractal spacetime structure in asymptotically safe gravity},
  JHEP {\bf 0510},(2005) 050 
\mbox{[{\tt{hep-th/0508202}}]}.
%
\bibitem{jan1}
M.~Reuter and J.-M.~Schwindt, \emph{A minimal length from the cutoff modes in asymptotically safe quantum
  gravity},
  JHEP {\bf 0601},(2006) 070 
\mbox{[{\tt{hep-th/0511021}}]}.
%
\bibitem{jan2}
M.~Reuter and J.-M.~Schwindt, \emph{Scale-dependent metric and causal structures in quantum Einstein gravity},
  JHEP {\bf 0701},(2007) 049 
\mbox{[{\tt{hep-th/0611294}}]}.
%
\bibitem{creh1}
M.~Reuter and H.~Weyer, \emph{Background Independence and Asymptotic Safety in Conformally Reduced Gravity},
Phys.\ Rev.\  D {\bf 79},(2009) 105005, 
\mbox{[{\tt{0801.3287 [hep-th]}}]}.
%
\bibitem{creh2}
M.~Reuter and H.~Weyer, \emph{Conformal sector of Quantum Einstein Gravity in the local potential
  approximation: non-Gaussian fixed point and a phase of \\
diffeomorphism invariance}, Phys.\ Rev.\  D in press,
\mbox{[{\tt{0804.1475 [hep-th]}}]}.
%
\bibitem{crehroberto}
P.~F.~Machado and R.~Percacci,
\emph{Conformally reduced quantum gravity revisited}, \mbox{[\tt 0904.2510 [hep-th]]}.
% 
\bibitem{elisa1}
  E.~Manrique and M.~Reuter,
  \emph{Bare Action and Regularized Functional Integral of Asymptotically Safe
  Quantum Gravity},
  Phys.\ Rev.\  D {\bf 79} (2009) 025008
 \mbox{[{\tt 0811.3888 [hep-th]}]}.
%
%
\bibitem{je1}
J.-E.~Daum and M.~Reuter, \emph{Effective Potential of the Conformal Factor: Gravitational Average Action
  and Dynamical Triangulations},
\mbox{[{\tt{0806.3907 [hep-th]}}]}.
%
\bibitem{max}
P.~Forg\'acs and M.~Niedermaier, \emph{A fixed point for truncated quantum Einstein gravity},
\mbox{[{\tt hep-th/0207028 }]};\\
M.~Niedermaier, \emph{On the renormalization of truncated quantum Einstein gravity},
  JHEP {\bf 0212} (2002) 066 \mbox{[{\tt hep-th/0207143}]};
\emph{Dimensionally reduced gravity theories are asymptotically safe},
  Nucl.\ Phys.\  B {\bf 673},(2003) 131, \mbox{[{\tt hep-th/0304117 }]};
\emph{The asymptotic safety scenario in quantum gravity: An introduction},
  Class.\ Quant.\ Grav.\  {\bf 24},(2007) R171 
\mbox{[{\tt gr-qc/0610018 }]}
%
\bibitem{livrev}
For detailed reviews of asymptotic safety in gravity see:\\
M.~Niedermaier and M.~Reuter,
\emph{The Asymptotic Safety Scenario in Quantum Gravity},
  Living Rev.\ Rel.\  {\bf 9},(2006) 5;\\ 
R.~Percacci, \emph{Asymptotic Safety}, \mbox{[{\tt{0709.3851 [hep-th].}}]}
%
\bibitem{avact}
C.~Wetterich, \emph{Exact evolution equation for the effective potential},
  Phys.\ Lett.\  B {\bf 301},(1993) 90.
%
\bibitem{nonabavact}
M.~Reuter, C.~Wetterich, \emph{Effective average action for gauge theories and exact evolution
  equations},
  Nucl.\ Phys.\  B {\bf 417},(1994) 181.
%
\bibitem{ym}
M.~Reuter and C.~Wetterich, 
\emph{Exact evolution equation for scalar electrodynamics},
  Nucl.\ Phys.\  B {\bf 427},(1994) 291;
\emph{Average action for the Higgs model with Abelian gauge symmetry},
  Nucl.\ Phys.\  B {\bf 391},(1993) 147; 
\emph{Running gauge coupling in three-dimensions and the electroweak phase
  transition},
  Nucl.\ Phys.\  B {\bf 408},(1993) 91,\\
M.~Reuter, \emph{Effective average action of Chern-Simons field theory},
  Phys.\ Rev.\  D {\bf 53} (1996) 4430
 \mbox{[{\tt hep-th/9511128}]};
 \emph{Renormalization of the Topological Charge in Yang-Mills Theory},
  Mod.\ Phys.\ Lett.\  A {\bf 12} (1997) 2777
 \mbox{[{\tt hep-th/9604124}]}.
%
\bibitem{liouv}
M.~Reuter and C. Wetterich,
\emph{Quantum Liouville field theory as solution of a flow equation},
  Nucl.\ Phys.\  B {\bf 506} (1997) 483
 \mbox{[{\tt hep-th/9605039}]}.
% [arXiv:hep-th/9605039].


\bibitem{avactrev}
J.~Berges, N.~Tetradis and C.~Wetterich,
\emph{Non-perturbative renormalization flow in quantum field theory and
  statistical physics},
  Phys.\ Rept.\  {\bf 363} (2002) 223
 \mbox{[{\tt hep-th/0005122}]}.\\
C.~Wetterich,
\emph{Effective average action in statistical physics and quantum field theory},
  Int.\ J.\ Mod.\ Phys.\  A {\bf 16},(2001) 1951 
  \mbox{[{\tt hep-ph/0101178 }]}.
%
\bibitem{ymrev}
For reviews of the effective average action in Yang--Mills theory see:\\
M.~Reuter, \emph{Effective Average Actions and Nonperturbative Evolution Equations},\mbox{[{\tt{hep-th/9602012}}]};
J.~Pawlowski, 
\emph{Aspects of the functional renormalisation group},
  Annals Phys.\  {\bf 322} (2007) 2831
 \mbox{[{\tt{hep-th/0512261}}]};\\
H.~Gies, \emph{Introduction to the functional RG and applications to gauge theories}, \mbox{[{\tt{hep-ph/0611146}}]}.
%
\bibitem{bh}
A.~Bonanno and M.~Reuter, \emph{}
\emph{Renormalization group improved black hole spacetimes},
 Phys.\ Rev.\ D {\bf 62} (2000) 043008, \mbox{[{\tt{hep-th/0002196}}]};
\emph{Spacetime structure of an evaporating black hole in quantum gravity}, 
Phys.\ Rev.\ D {\bf 73} (2006) 083005, \mbox{[{\tt{hep-th/0602159}}]};\\
\emph{Quantum gravity effects near the null black hole singularity}, 
Phys.\ Rev.\ D {\bf 60}  (1999) 084011, \mbox{[{\tt{gr-qc/9811026}}]}.
%
\bibitem{erick1}
M.~Reuter and E.~Tuiran, \emph{Quantum Gravity Effects in Rotating Black Holes},
 in Proceedings of the Eleventh Marcel Grossmann Meeting, H.Kleinert, R. Janzten, R. Ruffini (Eds),
World Scientific, Singapore (2007),
\mbox{[{\tt{hep-th/0612037}}]}.
%
\bibitem{cosmo1}
A.~Bonanno and M.~Reuter,
\emph{Cosmology of the Planck era from a renormalization group for quantum
  gravity},
  Phys.\ Rev.\  D {\bf 65},(2002) 043508 
  \mbox{[{\tt hep-th/0106133}]};\\
M.~Reuter and F.~Saueressig, 
\emph{From big bang to asymptotic de Sitter: Complete cosmologies in a  quantum
  gravity framework},
  JCAP {\bf 0509},(2005) 012 
\mbox{[{\tt hep-th/0507167}]}.
%
\bibitem{cosmo2}
A.~Bonanno and M.~Reuter,
\emph{Cosmology with self-adjusting vacuum energy density from a  renormalization
  group fixed point},
  Phys.\ Lett.\  B {\bf 527},(2002) 9 
\mbox{[{\tt{astro-ph/0106468}}]}; 
\emph{Cosmological perturbations in renormalization group derived cosmologies},
  Int.\ J.\ Mod.\ Phys.\  D {\bf 13} (2004) 107
\mbox{[{\tt{astro-ph/0210472}}]};\\
E.~Bentivegna, A.~Bonanno and M.~Reuter,
\emph{Confronting the IR Fixed Point Cosmology with High Redshift Supernova
  Data},
  JCAP {\bf 0401}, 001 (2004)
\mbox{[{\tt{astro-ph/0303150}}]}.
%
\bibitem{entropy}
A.~Bonanno and M.~Reuter,
\emph{Entropy signature of the running cosmological constant}
JCAP {\bf 0708}, (2007) 024 and \mbox{[{\tt arXiv:0706.0174 [hep-th]}]}.
%
\bibitem{esposito}
A.~Bonanno, G.~Esposito and C.~Rubano,
\emph{A class of renormalization group invariant scalar field cosmologies},
  Gen.\ Rel.\ Grav.\  {\bf 35},(2003) 1899 
 \mbox{[{\tt{hep-th/0303154}}]};
\emph{Arnowitt-Deser-Misner gravity with variable G and Lambda and fixed  point
  cosmologies from the renormalization group},
  Class.\ Quant.\ Grav.\  {\bf 21},(2004) 5005
\mbox{[{\tt{gr-qc/0403115}}]};
 A.~Bonanno, G.~Esposito, C.~Rubano and P.~Scudellaro,
\emph{The accelerated expansion of the universe as a crossover phenomenon},
  Class.\ Quant.\ Grav.\  {\bf 23},(2006) 3103 
\mbox{[{\tt{astro-ph/0507670}}]};
\emph{Noether symmetry approach in pure gravity with variable G and Lambda},
  Class.\ Quant.\ Grav.\  {\bf 24},(2007) 1443 
 \mbox{[{\tt{gr-qc/0610012}}]}.
%
\bibitem{h1}
M.~Reuter and H.~Weyer, 
\emph{Renormalization group improved gravitational actions: A Brans-Dicke
  approach},
  Phys.\ Rev.\  D {\bf 69},(2004) 104022 
\mbox{[{\tt{hep-th/0311196}}]}.
%
\bibitem{h2}
M.~Reuter and H.~Weyer,
\emph{Running Newton constant, improved gravitational actions, and galaxy
  rotation curves},
Phys.\ Rev.\ D {\bf 70} (2004) 124028
 \mbox{[{\tt{hep-th/0410117}}]}.
%
\bibitem{h3}
M.~Reuter and H.~Weyer,
\emph{Quantum gravity at astrophysical distances?},
JCAP {\bf 0412},  (2004) 001 
 \mbox{[{\tt{hep-th/0410119}}]}.
%
\bibitem{applecarra}
T.~Appelquist and J.~Carazzone,
\emph{Infrared Singularities And Massive Fields},
  Phys.\ Rev.\  D {\bf 11} (1975) 2856. 
%
\bibitem{girelli}
F.~Girelli, S.~Liberati, R.~Percacci and C.~Rahmede,
\emph{Modified dispersion relations from the renormalization group of  gravity},
  Class.\ Quant.\ Grav.\  {\bf 24},(2007) 3995. 
%
\bibitem{lhc}
D.~Litim and T.~Plehn,
\emph{Signatures of gravitational fixed points at the LHC},
  Phys.\ Rev.\ Lett.\  {\bf 100},(2008) 131301 
 \mbox{[{\tt 0707.3983 [hep-ph] }]}.
%
\bibitem{mof}
J.~Moffat,
\emph{Gravitational Theory, Galaxy Rotation Curves and Cosmology without Dark
  Matter},
  JCAP {\bf 0505},(2005) 003 
\mbox{[{\tt astro-ph/0412195 }]};\\
J.R.~Brownstein and J.~Moffat,
\emph{Galaxy Rotation Curves Without Non-Baryonic Dark Matter},
  Astrophys.\ J.\  {\bf 636},(2006) 721 
 \mbox{[{\tt astro-ph/0506370 }]};\\
\emph{Galaxy Cluster Masses Without Non-Baryonic Dark Matter},
  Mon.\ Not.\ Roy.\ Astron.\ Soc.\  {\bf 367},(2006) 527 
\mbox{[{\tt astro-ph/0507222  }]}.
%
\bibitem{back}
L.F.~Abbott,
\emph{Introduction To The Background Field Method},
 Acta Phys.\ Polon.\  B {\bf 13} (1982) 33; 
\emph{The Background Field Method Beyond One Loop},
  Nucl.\ Phys.\  B {\bf 185} (1981) 189;\\
B.S.~DeWitt, 
\emph{Quantum theory of gravity. II. The manifestly covariant theory},
  Phys.\ Rev.\  {\bf 162} (1967) 1195 ;\\
M.T.~Grisaru, P.van Nieuwenhuizen and C.C.~Wu,
\emph{Background Field Method Versus Normal Field Theory In Explicit Examples:
  One Loop Divergences In S Matrix And Green's Functions For Yang-Mills And
  Gravitational Fields},
  Phys.\ Rev.\  D {\bf 12} (1975) 3203;\\
D.M.~Capper, J.J.~Dulwich and M.\ Ramon Medrano,
\emph{The Background Field Method For Quantum Gravity At Two Loops},
  Nucl.\ Phys.\  B {\bf 254},(1985) 737;\\
S.L.~Adler, 
\emph{Einstein Graviy as a symmetry breaking effect in Quantum Field Theory},
  Rev.\ Mod.\ Phys.\  {\bf 54} (1982) 729
  [Erratum-ibid.\  {\bf 55} (1983) 837].
%
\bibitem{floper}
R.~Floreanini and R.~Percacci,
\emph{Average effective potential for the conformal factor},
  Nucl.\ Phys.\  B {\bf 436},(1995) 141 
 \mbox{[{\tt hep-th/9305172 }]};
\emph{Mean Field Quantum Gravity},
  Phys.\ Rev.\  D {\bf 46},(1992) 1566 
%
%
\bibitem{opt}
D.~Litim,
\emph{Optimisation of the exact renormalisation group},
  Phys.\ Lett.\  B {\bf 486},(2000) 92 
\mbox{[{\tt hep-th/0005245 }]};
\emph{Optimised renormalisation group flows},
  Phys.\ Rev.\  D {\bf 64},(2001) 105007 
\mbox{[{\tt hep-th/0103195 }]};
\emph{Mind the gap},
  Int.\ J.\ Mod.\ Phys.\  A {\bf 16},(2001) 2081 
\mbox{[{\tt hep-th/0104221}]}.
%
\bibitem{hamber}
H.W.~Hamber, 
\emph{Phases Of 4-D Simplicial Quantum Gravity},
  Phys.\ Rev.\  D {\bf 45},(1992) 507 ;
\emph{On the gravitational scaling dimensions},
  Phys.\ Rev.\  D {\bf 61} (2000) 124008
  \mbox{[{\tt hep-th/9912246 }]};
\emph{Discrete and Continuum Quantum Gravity},
\mbox{[{\tt 0704.2895 [hep-th]}]};\\
T.~Regge and R.M.~Williams, 
\emph{Discrete structures in gravity},
  J.\ Math.\ Phys.\  {\bf 41} (2000) 3964
\mbox{[{\tt gr-qc/0012035}]}.
%
\bibitem{ajl1}
J.~Ambj\o{}rn, J.~Jurkiewicz and R.~Loll,
\emph{Emergence of a 4D world from causal quantum gravity},
  Phys.\ Rev.\ Lett.\  {\bf 93} (2004) 131301
 \mbox{[{\tt hep-th/0404156}]}
%
\bibitem{ajl2}
J.~Ambj\o{}rn, J.~Jurkiewicz and R.~Loll,
\emph{Semiclassical universe from first principles},
  Phys.\ Lett.\  B {\bf 607} (2005) 205
 \mbox{[{\tt hep-th/0411152}]}
%
\bibitem{ajl3}
J.~Ambj\o{}rn, J.~Jurkiewicz and R.~Loll,
\emph{Spectral dimension of the universe},
  Phys.\ Rev.\ Lett.\  {\bf 95} (2005) 171301
 \mbox{[{\tt hep-th/0505113}]};
\emph{Reconstructing the universe},
  Phys.\ Rev.\  D {\bf 72} (2005) 064014
\mbox{[{\tt hep-th/0505154}]};
\emph{The universe from scratch},
  Contemp.\ Phys.\  {\bf 47},(2006) 103 
  \mbox{[{\tt hep-th/0509010}].}
%
\bibitem{zapcor}
E.~Manrique, R.~Oeckl, A.~Weber and J.~A.~Zapata,
\emph{Loop quantization as a continuum limit},
  Class.\ Quant.\ Grav.\  {\bf 23} (2006) 3393 
\mbox{[{\tt{hep-th/0511222}}]};
A.~Corichi, T.~Vukasinac and J.~A.~Zapata,
  \emph{Hamiltonian and physical Hilbert space in polymer quantum mechanics},
  Class.\ Quant.\ Grav.\  {\bf 24} (2007) 1495
\mbox{[{\tt{gr-qc/0610072}}]}.
%
\bibitem{giesymfp}
H.~Gies,
\emph{Renormalizability of gauge theories in extra dimensions},
 Phys.\ Rev.\  D {\bf 68} (2003) 085015
  \mbox{[{\tt hep-th/0305208}]}.
%
\bibitem{elisa2} E.~Manrique, M.~Reuter, work in progress.
%
\bibitem{nonlinsig}
A.~Codello and R.~Percacci,
\emph{Fixed Points of Nonlinear Sigma Models in} $d>2$,
  Phys.\ Lett.\  B {\bf 672} (2009) 280
\mbox{[{\tt0810.0715 [hep-th]}]}.
%

  %


\end{thebibliography}
\end{document}